\documentclass[longauth]{aa}

\usepackage{graphicx}
\usepackage{lscape}
\usepackage{longtable}
\usepackage{natbib}
\usepackage{color}
\usepackage{array} 
\usepackage{tikz,array}
\usetikzlibrary{calc}
\usepackage{multirow}
\usepackage{here}
\usepackage{txfonts}
\usepackage{amsmath,amstext}
\usepackage{epstopdf}
\usepackage{float}
\usepackage{mathtools}
\usepackage{booktabs}
\usepackage{subfigure}
\usepackage{url}
\usepackage{helvet}
\usepackage{tabularx}
\usepackage{multirow}
\usepackage{natbib}
\usepackage[flushleft]{threeparttable}
\usepackage{lscape}
\usepackage{pdflscape}
\usepackage{longtable}
\usepackage{wasysym}
\usepackage{float}
\usepackage{relsize}
\usepackage{color}
\usepackage{placeins}
\usepackage{breqn}
\usepackage{bm}
\usepackage{enumitem}
\usepackage{makecell}
\usepackage[colorlinks, citecolor=blue, linkcolor=blue]{hyperref}
\hypersetup{colorlinks,breaklinks, linkcolor=blue,urlcolor=magenta, anchorcolor=blue,citecolor=blue}
\bibpunct{(}{)}{;}{a}{}{,} % to follow the A&A style

\newcommand{\be}{\begin{equation}}
\newcommand{\ee}{\end{equation}}

\def\gcm3{\hbox{g cm$^{-3}$}}       %g.cm-3
\def\Msun{\hbox{$\mathrm{M}_{\astrosun}$}}             %Msun
\def\Rsun{\hbox{$\mathrm{R}_{\astrosun}$}}
\def\Mjup{\hbox{$\mathrm{M}_{\rm Jup}$}}
\def\Rjup{\hbox{$\mathrm{R}_{\rm Jup}$}}
\def\Mearth{\hbox{$\mathrm{M}_{\oplus}$}}
\def\Rearth{\hbox{$\mathrm{R}_{\oplus}$}}

\newcommand{\rebeal}{\texttt{rebeal}}

\newcommand{\gaia}{\textit{Gaia}}
\newcommand{\kepler}{\textit{Kepler}}
\newcommand{\cafe}{CAFE$_{\rm 2}$}

\def\gcm3{\hbox{g cm$^{-3}$}}       %g.cm-3
\def\Msun{\hbox{$\mathrm{M}_{\astrosun}$}}             %Msun
\def\Rsun{\hbox{$\mathrm{R}_{\astrosun}$}}
\def\Mjup{\hbox{$\mathrm{M}_{\rm Jup}$}}
\def\Rjup{\hbox{$\mathrm{R}_{\rm Jup}$}}
\def\Mearth{\hbox{$\mathrm{M}_{\oplus}$}}
\def\Rearth{\hbox{$\mathrm{R}_{\oplus}$}}

\usepackage{stfloats}

\usepackage{graphicx}
\usepackage{txfonts}
\begin{document}

% ===============================================================
% TITLES
% ===============================================================

   \title{CAFE follow-up of TESS hot Jupiter candidates left behind: \\ I. {Five} newly confirmed planets and a false positive\thanks{Based on observations collected at Centro Astron\'omico Hispano en Andaluc\'ia (CAHA) at Calar Alto, proposals 25A-2.2-014, 25B-2.2-014, 26A-2.2-015, operated jointly by Junta de Andaluc\'ia, and Consejo Superior de Investigaciones Cient\'ificas (IAA-CSIC).}}

	\titlerunning{\cafe{} follow-up of TESS hot Jupiter candidates left behind}
	\authorrunning{Lillo-Box et al.}

% ===============================================================
% AUTHORS
% ===============================================================

   \author{
%%--------------Lead Authors-----------------------
J.~Lillo-Box\inst{\ref{cab}} %\orcidicon{0000-0003-3742-1987}, }
%
% CAB:
\and C.~Cifuentes\inst{\ref{cab}}
\and O.~Balsalobre-Ruza\inst{\ref{cab}}
\and B.~Montesinos\inst{\ref{cab}}
% TESS SG coordinators
\and D.~Latham\inst{\ref{cfa}} 
\and K.~A.~Collins\inst{\ref{cfa}}   %0000-0001-6588-9574
\and D.~Ciardi\inst{\ref{nexsci}}  % Palomar/PHARO
%
% SOPHIE
\and G.~H\'ebrard\inst{\ref{IAP},\ref{ohp}} 
% NEID
\and S.~W.~Yee\inst{\ref{51peg},\ref{ucla}} 
% TLS
\and E.~W. Guenther\inst{\ref{tautenburg}} % {0000-0002-9130-6747} % TLS observations
% SOPHIE
\and H.~Bouy\inst{\ref{bordeaux},\ref{IUF}}  % SOPHIE
%
% Spell check
\and J.~N.~Winn\inst{\ref{princeton}} 
% === HR IMAGING ===
\and S.~B.~Howell\inst{\ref{ames}} % --> High-res imaging # [0000-0002-2532-2853]{} 
\and C.~Ziegler\inst{\ref{soar}}
\and M.~E.~Everett\inst{\ref{noirlab}} % High.Res. imaging # 0000-0002-0885-7215
\and B.~Safonov\inst{\ref{Sternberg}}
%
% === Ground-based observations ====
\and F.~Murgas\inst{\ref{iac},\ref{ull}} % MuSCAT2
\and N.~Narita\inst{\ref{komaba},\ref{astrobio},\ref{iac}} % MuSCAT2
%
% === HARPS for TOI-603
\and L.\,D.\,Nielsen\inst{\ref{munich},\ref{ull}} % MuSCAT2
\and A.~Abreu\inst{\ref{atg},\ref{ucm}} 
\and {J.~Aceituno}\inst{\ref{caha}}
\and {J.~F.~Ag\"u\'i~Fern\'andez}\inst{\ref{caha}}
\and {M.~Azzaro}\inst{\ref{caha}}
\and D.~Barrado\inst{\ref{cab}} 
\and P.~Benni\inst{\ref{acton}}
\and A.~Bieryla\inst{\ref{cfa}}
\and A.~Chontos\inst{\ref{dartmouth}}
\and C.~A.~Clark\inst{\ref{nexsci}} % High.Res. imaging # 0000-0002-2361-5812
\and E.~Delgado-Mena\inst{\ref{cab}} 
\and S.~J.~Deveny\inst{\ref{moffet},\ref{ames}} % --> High-res imaging # 0009-0002-9833-0667 
\and R.~For\'es-Toribio\inst{\ref{ohiouni},\ref{ohiocca}}
\and {J.~Flores-Mart\'in}\inst{\ref{caha}}
\and A.~Garmash\inst{\ref{novosibirsk}}
\and {S.~G\'ongora}\inst{\ref{caha}}
\and J.~D.Hartman\inst{\ref{princeton}}
\and C.~Haukes\inst{\ref{cab}} 
\and N.~Heidari\inst{\ref{IAP}} 
\and I.~J.~Helm\inst{\ref{mason}}
\and J.~F.~Kielkopf\inst{\ref{louisville}}
\and J.~K\"ohler\inst{\ref{tautenburg}} % TLS observations
\and C.~Littlefield\inst{\ref{moffet},\ref{ames}} % --> High-res imaging # 0000-0001-7746-5795 
\and M.~B.~Lund\inst{\ref{nexsci}} % High.Res. imaging # 0000-0003-2527-1598
\and B.~Massey\inst{\ref{landers}}
\and A.~Masson\inst{\ref{cab}} 
\and J.~McCormac\inst{\ref{warwick}} 
\and M.~Morales-Calder\'on\inst{\ref{cab}} 
\and M.~Mori\inst{\ref{astrobio},\ref{naoj}} % MuSCAT2
%\and C.~Moutou\inst{\ref{toulouse}} 
\and J.~A.~Mu\~noz\inst{\ref{uv},\ref{oauv}}
\and E.~Pall\'e\inst{\ref{iac},\ref{ull}} % MuSCAT2
\and P.~Plavchan\inst{\ref{mason}}
\and M.~Reefe\inst{\ref{mit}}
\and J.~E.~Schlieder\inst{\ref{goddard}} % High.Res. imaging
\and G.~Srdoc\inst{\ref{Croatia}}
\and C.~Stockdale\inst{\ref{hazelwood}}
\and J.~C.~van~Eyken\inst{\ref{nexsci}} % High.Res. imaging # 0000-0003-2192-5371
\and R.~G.~West\inst{\ref{warwick}} 
\and F.~P.~Wilkin\inst{\ref{union}} 
\and J.~Wittrock\inst{\ref{missouri}}
\and D.-V.~Zora\inst{\ref{union}}
% \inst{\ref{}}
%------------At least Read & Comment -------
%
}

% ===============================================================
% AFFILIATIONS
% ===============================================================

\institute{
% ------------------------------------------
Centro de Astrobiolog\'ia (CAB), CSIC-INTA, Camino Bajo del Castillo s/n, 28692, Villanueva de la Ca\~nada (Madrid), Spain \email{Jorge.Lillo@cab.inta-csic.es} 
\label{cab}
% ------------------------------------------
\and
Center for Astrophysics \textbar \ Harvard \& Smithsonian, 60 Garden Street, Cambridge, MA 02138, USA 
\label{cfa}
% ------------------------------------------
\and
NASA Exoplanet Science Institute-Caltech/IPAC, Pasadena, CA 91125, USA
\label{nexsci}
% ------------------------------------------
\and
Institut d'astrophysique de Paris, UMR7095 CNRS, Universit\'e Pierre \& Marie Curie, 98bis boulevard Arago, 75014 Paris, France
\label{IAP}
% ------------------------------------------
\and
Observatoire de Haute-Provence, 04670 Saint Michel l'Observatoire, France
\label{ohp}
% ------------------------------------------
\and
51 Pegasi b Fellow
\label{51peg}
% ------------------------------------------
\and
Department of Physics \& Astronomy, University of California Los Angeles, Los Angeles, CA 90095, USA
\label{ucla}
% ------------------------------------------
\and
Th\"uringer Landessternwarte Tautenburg, 07778 Tautenburg, Germany 
\label{tautenburg}
% ------------------------------------------
\and
Laboratoire d'astrophysique de Bordeaux, Univ. Bordeaux, CNRS, B18N, all\'ee Geoffroy Saint-Hilaire, 33615 Pessac, France.
\label{bordeaux}
% ------------------------------------------
\and
Institut universitaire de France (IUF), 1 rue Descartes, 75231 Paris CEDEX 05
\label{IUF}
% ------------------------------------------
\and
Department of Astrophysical Sciences, Princeton University, 4 Ivy Lane, Princeton, NJ 08544, USA
\label{princeton}
% ------------------------------------------
\and
NASA Ames Research Center, Moffett Field, CA 94035, USA
\label{ames}
% ------------------------------------------
\and
Department of Physics, Engineering and Astronomy, Stephen F. Austin State University, 1936 North St, Nacogdoches, TX 75962, USA
\label{soar}
% ------------------------------------------
\and
NSF NOIRLab, 950 N. Cherry Ave., Tucson, AZ 85719, USA
\label{noirlab}
% ------------------------------------------
\and
Sternberg Astronomical Institute Lomonosov Moscow State University, Universitetskii prospekt, 13, 119992, Moscow, Russia.
\label{Sternberg}
% ------------------------------------------
\and
Instituto de Astrof\'{i}sica de Canarias (IAC), 38205 La Laguna, Tenerife, Spain
\label{iac}
% ------------------------------------------
\and
Departamento de Astrof\'isica, Universidad de La Laguna (ULL), E-38206 La Laguna, Tenerife, Spain
\label{ull}
% ------------------------------------------
\and
Komaba Institute for Science, The University of Tokyo, 3-8-1 Komaba, Meguro, Tokyo 153-8902, Japan
\label{komaba}
% ------------------------------------------
\and
Astrobiology Center, 2-21-1 Osawa, Mitaka, Tokyo 181-8588, Japan
\label{astrobio}
% ------------------------------------------
\and 
University Observatory, Faculty of Physics, Ludwig-Maximilians-Universit{\"a}t M{\"u}nchen, Scheinerstr. 1, 81679 Munich, Germany
\label{munich}
% ------------------------------------------
\and
ATG Science \& Engineering for the European Space Agency (ESA), ESAC, Spain
\label{atg}
% ------------------------------------------
\and
Universidad Complutense de Madrid, Av. Complutense, s/n, Moncloa - Aravaca, 28040 (Madrid), Spain
\label{ucm}
% ------------------------------------------
\and
Centro Astron\'omico Hispano en Andaluc\'ia, Observatorio de Calar Alto, Sierra de los Filabres, E-04550 G\'ergal, Almer\'ia, Spain
\label{caha}
% ------------------------------------------
\and
Acton Sky Portal private observatory, Acton, MA, USA
\label{acton}
% ------------------------------------------
\and
Department of Physics and Astronomy, Dartmouth College, Hanover, NH 03755, USA
\label{dartmouth}
% ------------------------------------------
\and
Bay Area Environmental Research Institute, Moffett Field, CA 94035, USA
\label{moffet}
% ------------------------------------------
\and
Department of Astronomy, The Ohio State University, 140 West 18th Avenue, Columbus, OH 43210, USA
\label{ohiouni}
% ------------------------------------------
\and
Center for Cosmology and Astroparticle Physics, The Ohio State University, 191 W. Woodruff Avenue, Columbus, OH 43210, USA
\label{ohiocca}
% ------------------------------------------
\and
Physics and Astronomy Laboratory, Lyceum 130, 10 Uchenyh St., Novosibirsk, 630090, Russia
\label{novosibirsk}
% ------------------------------------------
\and
Villa '39 Observatory, Landers, CA 92285, USA
\label{landers}
% ------------------------------------------
\and
Dept.\ of Physics, University of Warwick, Gibbet Hill Road, Coventry CV4 7AL, UK
\label{warwick}
% ------------------------------------------
\and
National Astronomical Observatory of Japan, 2-21-1 Osawa, Mitaka,
Tokyo 181-8588, Japan
\label{naoj}
% ------------------------------------------
%\and
%Universit\'e de Toulouse, CNRS, IRAP, 14 avenue Belin, 31400 Toulouse, France
%\label{toulouse}
% ------------------------------------------
\and
Departamento de Astronom\'{\i}a y Astrof\'{\i}sica, Universidad de Valencia, E-46100 Burjassot, Valencia, Spain
\label{uv}
% ------------------------------------------
\and
Observatorio Astron\'omico, Universidad de Valencia, E-46980 Paterna, Valencia, Spain
\label{oauv}
% ------------------------------------------
\and
George Mason University, 4400 University Dr, Fairfax, VA 22030
\label{mason}
% ------------------------------------------
\and
Department of Physics and Astronomy, University of Louisville, Louisville, KY 40292, USA
\label{louisville}
% ------------------------------------------
\and
Kavli Institute for Astrophysics \& Space Research, Massachusetts Institute of Technology, Cambridge, MA02139, USA
\label{mit}
% ------------------------------------------
\and
NASA Goddard Space Flight Center, Greenbelt, MD 20771, USA
\label{goddard}
% ------------------------------------------
\and
Kotizarovci Observatory, Sarsoni 90, 51216 Viskovo, Croatia
\label{Croatia}
% ------------------------------------------
\and
Hazelwood Observatory, Australia
\label{hazelwood}
% ------------------------------------------
\and
Department of Physics and Astronomy, Union College, 807 Union St., Schenectady, NY 12308, USA
\label{union}
% ------------------------------------------
\and
Department of Physics, Astronomy, \& Materials Science, Missouri State University, Springfield, MO 65897, USA
\label{missouri}
% ------------------------------------------
}

% ===============================================================
% DATE
% ===============================================================

   \date{In prep.}

% ===============================================================
% ABSTRACT & KEYWORDS
% ===============================================================

% \abstract{}{}{}{}{}
% 5 {} token are mandatory
 
  \abstract
  % context heading (optional)
 % {} leave it empty if necessary  
   {Hot Jupiters are key targets for understanding planet formation, migration and atmospheres. Yet, most of the ground-based observational resources for the follow-up of the TESS mission are focused on the confirmation of low-mass planet candidates. This leaves the more massive planets, such as hot Jupiters, without mass determinations and definitive confirmations.}
  % aims heading (mandatory)
   {We use the \cafe{} spectrograph at the Calar Alto Observatory to monitor the radial velocity of stars with hot-Jupiter candidates that have been left behind by follow-up efforts with the goal of confirming their planetary nature. Here, we present the results for seven candidates.}
  % methods heading (mandatory)
   {We monitor the radial velocity of the seven candidate host stars TOI-603, TOI-1137, TOI-1837, TOI-2114, TOI-4492, TOI-5806, and TOI-5811. Each hosts a hot-Jupiter candidate detected by the TESS mission. We characterize these stars and jointly model the \cafe{} radial velocities and the TESS photometric data to unravel the nature of the transiting objects and to infer their properties.}
  % results heading (mandatory)
   {The objects confirmed to be new planets are TOI-603\,b (mass $33.0^{+6.5}_{-6.2}$~\Mearth, period $16.2$~d), TOI-2114\,b ($1.01^{+0.14}_{-0.12}$~\Mjup, $6.2$~d), TOI-4492\,b ($5.92^{+0.67}_{-0.64}$~\Mjup, $4.4$~d), TOI-5806\,b ($2.77^{+0.34}_{-0.32}$~\Mjup, $3.2$~d), {and TOI-5811\,B\,b ($0.81^{+0.11}_{-0.10}$~\Mjup, $6.3$~d)}. Based on their masses and periods, TOI-603\,b is a member of the ``Neptune savanna'', while TOI-2114\,b, TOI-4492\,b, TOI-5806\,b, {and TOI-5811\,B\,b} are hot Jupiters orbiting slightly evolved stars. {We found the candidate TOI-5811.01 to actually be a planet but orbiting the nearby bound companion TOI-5811\,B}. We also identify a bound stellar companion to TOI-5806 at a projected separation of 248~au, thus {both} forming S-type (satellite-type) planetary systems. As for the other {two} candidates, TOI-1837.01 is definitely an eclipsing binary, while the nature of TOI-1137.01 remains indeterminate.}
  % conclusions heading (optional), leave it empty if necessary
   {The {five} newly confirmed planets orbit relatively bright stars (${\rm G}=8.6-10.2$~mag) and {four} of them are excellent candidates for atmospheric studies (with transmission spectroscopy metrics exceeding 90 for TOI-603\,b, TOI-2114\,b, TOI-5806\,b, {and TOI-5811\,B\,b}). The complex configurations of {all these} systems underscore the need for intensive follow-up efforts to clarify the nature of transiting planet candidates.} 


   \maketitle

% ===============================================================
% PAPER
% ===============================================================
\nolinenumbers
%

%===========================================================================
\section{Introduction}
%===========================================================================

Close-in massive planets were the first to be discovered in the exoplanet exploration era. Indeed, 51\,Peg\,b \citep{mayor95} belonged to this population of planets, and the subsequent detections increasingly populated this region of parameter space. However, one of the main conclusions from the first few decades of exoplanet hunting -- especially the \kepler{} mission \citep{borucki03} -- was the rarity of close-orbiting giant planets relative to smaller planets \citep{borucki11}, after correcting for observational biases. In the pre-\kepler{} era, estimated occurrence rates of hot-Jupiters ($R_{\rm p}>7~R_{\oplus}$, $P<10$~days) ranged from the $0.10^{+0.27}_{-0.08}$~\% \citep{bayliss11} using the transit technique to $1.2 \pm 0.38$~\% \citep{wright12} and $0.9 \pm 0.4$~\% \citep{mayor11} using radial velocity surveys. This low frequency was confirmed by using \kepler{} data, with which \cite{kunimoto20} inferred a hot-Jupiter occurrence rate of $0.77^{+0.16}_{-0.14}$~\%. All these results show overall that hot Jupiters occur rarely around Sun-like stars.

However, hot Jupiters are highly valuable. Over most of the past two decades, they have been the only accessible targets for atmospheric explorations. Thanks to their high irradiation, we have started to witness interesting chemical processes in their atmospheres (e.g., the photochemistry inferred by the detection of the SO$_2$ molecule in WASP-39\,b by \citealt{rustamkulov23}) and to unveil a plethora of atomic elements and molecules (see the review by \citealt{madhusudhan16}, and references therein). The existence of hot Jupiters led theorists to investigate mechanisms for orbital migration (e.g., \citealt{lin96}) and fed theoretical works on planet formation scenarios (see, e.g., \citealt{heller18}, and references therein). Their mere presence is associated with changes in the architecture of the rest of a planetary system by, for instance, lowering the probability of nearby low-mass companions (e.g., \citealt{knutson14,sha26}).

The Transiting Exoplanet Survey Satellite (TESS) mission \citep{ricker14} has produced thousands of new planetary candidates. Most of them have orbital periods less than a few weeks, as dictated by the mission's survey strategy (with fields of view being monitored continuously for only 27 days). As of today, among the more than 7700 planet candidates announced by the TESS mission science team, 3\,958 have orbital periods shorter than 10 days and inferred planetary radii larger than 7~\Rearth. Among this sample of candidate hot gas giants, 250 have already been identified as eclipsing binaries, 398 were previously known planets, and only 90 are new planets detected by TESS and subsequently confirmed by the community since the launch of the mission in 2018. This leaves about 3\,200 candidates ($\sim$80~\% of the original candidates) that still lack a mass measurement and thus a definitive confirmation and characterization to establish the nature of the transiting object. 

In this paper, we present the first results of a radial velocity follow-up effort with the \cafe{} (Calar Alto Fiber-fed \'Echelle spectrograph) instrument \citep{aceituno13,lillo-box19} to measure the dynamical masses of TESS hot-Jupiter candidates. In Sect.~\ref{sec:observations}, we present the target selection of our survey and describe our observational techniques. In Sect.~\ref{sec:host}, we analyse the properties of the stars in the sample and in Sect.~\ref{sec:analysis} we attempt to use all available data to verify or reject the hypothesis that each candidate is truly a transiting planet, and if so, to characterize the system parameters. In Sect.~\ref{sec:conclusions}, we conclude with some remarks and future prospects.

%===========================================================================
\section{Observations}
%===========================================================================
\label{sec:observations}

%++++++++++++++++++++++++++++++++++++++++++++++++++++
\subsection{The \cafe{}-TESS follow-up program: target selection}
%++++++++++++++++++++++++++++++++++++++++++++++++++++

The observations presented in this paper are the result of a Large Program (PI: J.~Lillo-Box) running on the 2.2-meter telescope at the Calar Alto (CAHA) observatory using the \cafe{} instrument \citep{aceituno13,lillo-box19}. Our target selection is focused on the targets accessible and with potential of being probed from such facility, which requires i) relatively bright stars (typically $G<12$~mag), ii) Northern hemisphere coordinates (${\rm DEC}>-10^{\circ}$), iii) planet candidates with radius larger than 7~\Rearth, iv) orbital periods shorter than 10 days, and v) labelled as "Planet Candidate" by the TESS Follow-up Observing Program (TFOP\footnote{\url{https://exofop.ipac.caltech.edu/tess/}}). Since the beginning of observations of our candidates, \cafe{} has already contributed to the confirmation of two of them, namely TOI-1408 \citep{korth24}, in which the hot-Jupiter TOI-1408\,b and an inner 2.2~\Mearth{} companion exist in a 2:1 mean motion resonance, and TOI-1471 \citep{osborn23}, a system of two warm Neptunes. In this paper, we present follow-up results for seven additional candidates: TOI-603.01, TOI-1137.01, TOI-1837.01, TOI-2114.01, TOI-4492.01, TOI-5806.01, and TOI-5811.01. Table~\ref{tab:Targets} summarizes the properties of the candidates reported in the TESS Project Candidates table\footnote{\url{https://exoplanetarchive.ipac.caltech.edu/cgi-bin/TblView/nph-tblView?app=ExoTbls&config=TOI}}. %In the process of obtaining the data presented in this paper, the candidate TOI-603.01 has been confirmed by \cite{chontos24} using HIRES data. Still, we also present the results for this target in this paper as an independent confirmation.   

%_____________________________________________________________
%
\begin{table*}
\caption{General properties of the targets studied in this paper from bibliography. We include the TESS Object of Interest (TOI) and TESS Input Catalog (TIC) names and coordinates {of the hosts identified by TESS} as well as the inferred preliminary properties of their planetary candidates from the SPOC {(Science Processing Operations Center; \citealt{jenkins16})} alert service. No uncertainties are reported here on P$_{\rm orb}$, T$_0$ and R$_p$ (those can be found in ExoFOP) as they are only reference values. }             % title of Table
 \setlength{\extrarowheight}{3pt}
 \small
\label{tab:Targets}      % is used to refer this table in the text
\centering                          % used for centering table
\begin{tabular}{r r r r r r r r r r l }        % centered columns (4 columns)
\toprule
TOI & TIC & RA (J2015.5) & DEC (J2015.5) & $T_{\rm mag}$ & P$_{\rm orb}$ &  R$_{\rm p}$ & T$_0$  & T$_{\rm dur}$ & $\delta$ & $d$    \\    % table heading
    &     & (deg)        & (deg.)        &  (mag)        & (days)        &  (\Rearth)   & (days) & (h)           & (ppt)    & (pc)           \\    % table heading
\midrule                     
603.01  & 262746281 & 141.10785421  & 5.76605623259  & 9.74 & 16.179860 & 7.90  & 3271.218286 & 6.5 & 2.249 & $204.26 \pm 0.60$\\
1137.01 & 89389197  & 301.177679821 & 30.4099493357  & 8.91 & 1.3206356 & 34.7  & 2770.240602 & 1.6 & 1.052 & $415.4 \pm 2.1$\\
1837.01 & 144441148 & 207.905911258 & 33.4923901244  & 8.71 & 5.8182238 & 7.82  & 1931.250769 & 1.2 & 1.070 & $152.7 \pm 3.0$\\
2114.01 & 9828416   & 261.096394621 & 33.2050726425  & 9.83 & 6.2099619 & 14.37 & 3476.661598 & 4.5 & 4.090 & $314.1 \pm 1.3$\\
4492.01 & 298297931 & 308.650800028 & 29.2381015582  & 9.63 & 4.4331838 & 12.52 & 3555.320385 & 3.9 & 9.166 & $142.82 \pm 0.32$\\
5806.01 & 283621618 & 329.101802875 & 20.6779808001  & 8.29 & 3.1856488 & 11.27 & 2850.812572 & 1.3 & 4.750 & $118.43 \pm 0.26$\\
5811.01 & 331484419 & 310.911906127 & 19.7279381189  & 7.75 & 6.25656   & 11.49 & 3549.411097 & 3.5 & 0.875 & $167.8 \pm 1.6$\\
\bottomrule
\end{tabular}
\end{table*}

%_____________________________________________________________

%+++++++++++++++++++++++++++++
\subsection{TESS photometry}
%+++++++++++++++++++++++++++++
\label{sec:TESS}
We obtained the TESS photometric time series for our seven targets through the \texttt{lightkurve}\footnote{\url{https://lightkurve.github.io/lightkurve/}} Python module \citep{lightkurve}. In the interest of homogeneity, and because the candidates studied in this paper generally produce deep transit signals, we only use light curves extracted with the mission pipeline SPOC (Science Processing Operations Center; \citealt{jenkins16}). Thus, we use the SPOC light curves when available; otherwise, the TESS-SPOC light curves {(obtained from the Full Frame images; \citealt{caldwell20})} are used. Even when data with 20\,s time sampling are
available, we still use the 120\,s SPOC light curve. 

We explored the field around the targets using the \texttt{tpfplotter}\footnote{Publicly available at \url{https://github.com/jlillo/tpfplotter}} \citep{aller20} tool. This Python module combines the \gaia{} DR3 catalog \citep{GaiaCollaboration2023} with the Target Pixel File (TPF) timestamps from the TESS observations to check for the possible contamination of nearby sources. The plots for each of the targets are shown in Figs.~\ref{fig:tpfplots}-\ref{fig:tpfplots2}. TOI-603, TOI-1137, TOI-4492 and TOI-5811 have relatively bright nearby sources outside of the apertures, and  TOI-4492, TOI-5806 and TOI-5811 also have close sources within the same pixel as the target star. On the other hand, TOI-1837 and TOI-2114 are very isolated targets, with negligible contamination from other stars. 

For stars with bright faraway sources, it is relevant to analyse whether these could be the origin of the transit signal. To this end, we use the TESS Positional Probability\footnote{\url{https://github.com/ahadjigeorghiou/TESSPositionalProbability}} code \citep{tpp}. This code analyses the probability that the target or any of the nearby sources is the cause of the detected transit, based on comparison of the observed centroid offset and models of such value for different potential false positive configurations. We run this code for all seven targets. The results are presented in Table~\ref{tab:TESSprob}. The outcome of this analysis shows that the  TESS photometric measurements for TOI-1137, TOI-1837 and TOI-5811 are strongly contaminated by nearby companions, leaving ambiguous the identity of the star showing transit-like dips. In the case of TOI-1837, there are three possible sources having a probability $>$1\% of being the origin of the transit. All of these sources are located between 1.5-3 pixels\footnote{TESS pixels are $21\arcsec \times 21\arcsec$ in sky-projected size.} away from the target and are 7 mag fainter than the target. We use the \texttt{tess-cont} module \citep{castro-gonzalez24b} to estimate by what fraction each source's flux would need to drop in order for the total light measured by TESS to exhibit the observed transit depth. For all three companions, the variation would need to exceed 100\%, i.e., none of them could be responsible for the transit-like signals. We conclude that the transit signals most likely originate from the intended target (TIC 144441148, TOI-1837). We proceed in a similar fashion with the other targets with companions that show probabilities of being the transit hosts above 1\% (see Table~\ref{tab:TESSprob}). 
For TOI-1137, the faint companion could only be the source of the transits if it were a binary with 82\% eclipse depths -- practically impossible -- but we nevertheless investigated this case by extracting a TESS lightcurve using TESScut\footnote{\url{https://mast.stsci.edu/tesscut/}}. The resulting photometric time series show strong variability but nothing approaching 82\%, nor with the 1.3-day periodicity found for the TOI-1137.01 candidate. Hence we can discard this other source as the origin of the transits. 
For TOI-5806, we find that the relatively bright ($\Delta G=3.6$~mag) and close ($\rho=2.1\arcsec$) source could actually be the origin of the transits with a 11.5\% probability. Indeed, the transits there would need to be 10\% (i.e., 100 ppt) deep to mimic the current transits on the target. When extracting the TESS light curve of this target, we also find the periodic transits with a contaminated depth of 4.5 parts per thousand (ppt). When correcting for the contamination of our target (assuming the transit occurs in this other star), the uncontaminated depth corresponds to 15 ppt. This is around an order of magnitude smaller than the required 100 ppt mentioned above and calculated by \texttt{tess-cont}, thus suggesting that this companion is not the source of the eclipses, and increasing the likelihood that the transits originate from the intended target. 
For TOI-5811, the only possible alternative {(resolved in Gaia DR3)} to the intended target is its visual binary companion, located at $7.6\arcsec$ separation and having a very similar magnitude ($\Delta G=0.7$~mag). Indeed, the alternative depth analysis shows that this target would need to undergo 0.17\% (1.7 ppt) transits in order to be the source of photometric variations observed by TESS. The photometric time series of this alternative star {actually show transits with such depth; consequently} we cannot rule out this possibility. 

Hence, from this analysis, the only ambiguous case is TOI-5811, where the companion TIC\,331484427 (HD\,197518B; TOI-5811\,B) might be the source of the transit signal rather than the targeted star. {This ambiguity is resolved in Sect.~\ref{sec:results-5811} using CARMENES and TRES RVs}

%_____________________________________________________________
%
\begin{table}
\caption{Probabilities of hosting the transit signal (P$_{\rm tr}$) for the main source and nearby sources (when P$_{\rm tr}>1$\%) estimated by the TESS Positional Probability code \citep{tpp} on stars identified in the \gaia{} DR3 catalog.}             % title of Table
 \setlength{\extrarowheight}{3pt}
 \small
\label{tab:TESSprob}      % is used to refer this table in the text
\centering                          % used for centering table
\begin{tabular}{r r r r r r}        % centered columns (4 columns)
\toprule               
TOI & P$_{\rm tr}$  & Alternatives (P$_{\rm tr}>1$\%) & $\rho$ & $\Delta G$ & Feas.$^{\dagger}$ \\    % table heading
    &   (\%)        & TIC (P$_{\rm tr}$)   &  ($^{\prime\prime}$) &  (mag)  & Y/N    \\    % table heading
\midrule                     
  603  & 100\% &  & & &  \\
  1137 & 91.2\% & 89389193 (7.9\%)  & 20.6 & +7.1 & N   \\
%  1719 & 100\% &  & & &    \\
  1837 & 33.9\% & 144441145 (29.4\%)  & 45.5 & +7.7 & N   \\
       &        & 144441147 (24.0\%)  & 30.7 & +7.5 & N \\
       &		& 144441146 (12.7\%)  & 58.4 & +7.1 & N \\
  2114 & 100\% &  & & &   \\
  4492 & 100\% &  & & &   \\ 
  5806 & 88.5\% & 2000790966 (11.5\%)  & 2.1 & +3.6 & N  \\
  5811 & 60.5\% & 331484427 (35.9\%)  & 7.6 & +0.7 & Y \\
  		&		& 331477375 (3.6\%)  & 135.6 & +1.9 & N  \\
\bottomrule
\end{tabular}
\tablefoot{
$^{(\dagger)}$ Feasibility of the nearby source for being the origin of the transit (see Sect.~\ref{sec:TESS}).
}
\end{table}

%_____________________________________________________________

%+++++++++++++++++++++++++++++++++++++++++
\subsection{High-resolution spectroscopy}
%++++++++++++++++++++++++++++++++++++++++++

All targets studied in this work were initiated with the \cafe{} follow-up Large Program, obtaining between 9 and 29 measurements per target. Additionally, data from other high-resolution spectrographs were obtained along the process. Here we describe the observations from all instruments capable of obtaining precise radial velocities. The time series for each target, together with the cross-correlation function (CCF) properties and applied drift and NZP corrections (when applicable) for each spectrum are shown in Table~\ref{tab:RVdata}.

%------------------------------
\subsubsection{CAFE$_2$}
%------------------------------
\label{sec:CAFEobs}
We used the Calar Alto Fiber-fed \'Echelle (\cafe{}) spectrograph, installed at the 2.2m telescope of the Calar Alto Observatory (Almer\'ia, Spain) to monitor our seven targets {(see Table~\ref{tab:RVdata} for a detailed description of the dataset per target)}. \cafe{} is located in an isolated room with active temperature control but is not stabilized in pressure. As a consequence, the science exposures need to be interspersed with exposures of a ThAr calibration source to monitor the nightly drift of the instrument. We also monitor night-to-night RV offsets by observing the same standard stars in all observing runs. The instrument provides a spectral resolution per {extracted} pixel of $R \sim 62\,000$ and a wavelength coverage of 400-950\,nm. The data are automatically processed by the on-site pipeline (described by \citealt{lillo-box19}), which also provides a RV estimation using the cross-correlation technique \citep{baranne96,pepe02}, with a G2 mask. We used the extracted s2d spectra to obtain our own RVs by cross-correlating them with masks from the closer spectral types to each of the targets. To this end, we used the \texttt{shaq} pipeline (see \citealt{lillo-box22}), which provides the radial velocity measured on the spectrum together with the full-width at half-maximum (FWHM) and the bisector (BIS) of the CCF, both indicators of the line shape. The RV value is then corrected by the drift, estimated at the mid-exposure time as a spline interpolation using the whole drift time series obtained along the same night through ThAr frames. This is also performed for the RV standards, which are then used to estimate the nightly zero points (NZPs) of the instrument by following similar prescriptions as explained in \cite{trifonov18}. 

For TOI-5811, we identified a second peak in the CCF, with around one third of the contrast of the primary peak, that appeared in the data from campaigns in 2025 but was not present in the 2024 data. This is in line with the \gaia{} non-single star catalog, which reports this source to be a spectroscopic binary with a period of $916 \pm 23$\,d and an eccentricity of $0.634 \pm 0.059$. We modeled the full CCF with two Gaussian profiles representing star A (deeper CCF) and star B (shallower CCF), and obtained the RVs of each component by determining the centroid position of each Gaussian component.

%------------------------------
\subsubsection{TRES}
%------------------------------
The Tillinghast Reflector \'Echelle Spectrograph (TRES; \citealt{gaborthesis}) is a fiber-fed moderately high-resolution spectrograph ($R \sim 44\,000$) in the optical band (390-910 nm) located at the Fred Lawrence Whipple Observatory (Arizona, US). 
We obtained observations of {six} of the targets, namely: TOI-603 (six visits), TOI-1137 (three), TOI-2114 (two), TOI-4492 (two), TOI-5811\,A (two), {and TOI-5811\,B (two)}. TRES observes over a wavelength range of 390-910 nm. Wavelength-calibrated and intensity rectified spectra were extracted using the standard procedures described in \cite{buchhave11}. The spectra of TOI-5811 were clearly double-lined. Multi-order relative RVs were derived by a correlation analysis against a master template created for each target from the observed spectra of that star. For the uncertainty in each observation we report the rms of the values found for the individual \'echelle orders included in the analysis, typically about 20 in number. Typical uncertainties for the RVs from these reconnaissance spectra are in the range of 20 to 30 m/s.

%------------------------------
\subsubsection{SOPHIE} 
\label{sec:SOPHIE}
%------------------------------
We observed the targets TOI-2114 and TOI-5806 also with the High-resolution mode of the SOPHIE spectrograph \citep{perruchot08} at the Observatoire de Haute-Provence (OHP), installed at the 1.93-meter telescope.  
For TOI-2114, a total of four spectra in three nights were gathered between 15-oct-2025 and 18-oct-2025 (PI: H. Bouy). 
For TOI-5806, we use 61 archival spectra obtained between 30-jul-2013 and 30-dec-2014 as part of the SuperWASP follow-up (PI: G. H\'ebrard). In both cases, the data were reprocessed with the current instrument pipeline, which also extracts the RVs and activity indicators such as the BIS. In the case of TOI-5806 which is a fast rotator, we applied the optimized procedure described in \cite{heidari22,heidari24}. For TOI-2114, the average precision of the four RV measurements is around 3 m/s, while 23 m/s is achieved for  TOI-5806. In this latter case, the time series have a large scatter corresponding to 200 m/s, thus suggesting the presence of astrophysical signals in it. The Generalized Lomb Scargle (GLS) periodogram of the RVs shows a peak at the $\sim 3.2$~d periodicity of the transiting signal. Although a moderate correlation is identified  between RVs and BIS, with Pearson's r of 0.41 (and p-value$=0.01$), no significant peak is clearly identified at the transiting period in the BIS. Instead, both datasets show some long-period signals at several hundred days, potentially indicating some long-term activity. This will be analysed together with the whole dataset in Sect.~\ref{sec:TOI-5806}. 

%------------------------------
\subsubsection{{HIRES}}
%------------------------------

{We observed TOI-603 with the High-Resolution Echelle Spectrometer (HIRES, \citealt{vogt94}) installed on the 10-m Keck I telescope on Maunakea in Hawai'i. HIRES is a grating, cross-dispersed, \'echelle spectrograph providing a resolution of 85\,000 in the 300-1000\,nm spectral regime. An iodine-cell is used for simultaneous wavelength calibration. A total of 12 observations of TOI-603 were obtained between 28-nov-2020 and 13-may-2022. The data were reduced using the well-tested procedures of the California Planet Search survey, described in \cite{howard10}. They produced an average RV uncertainty of 2.2 m/s.}

%------------------------------
\subsubsection{{HARPS}}
%------------------------------

{We used the High Accuracy Radial velocity Planet Searcher (HARPS, \citealt{mayor03}) at the European Southern Observatory (ESO) La Silla observatory to monitor the RV of TOI-603 in a time span of 50 days between 24-nov-2020 to 13-jan-2021. The HARPS spectrograph provides a resolution of $R=120\,000$ within the optical range 378-691\,nm. The nine spectra  were processed by the instrument Data Reduction Software (DRS) and the radial velocities were extracted through cross-correlation with a G2 mask, resulting in an average RV uncertainty of 2.6 m/s.}

%------------------------------
\subsubsection{{CARMENES}}
%------------------------------
\label{sec:CARMENES}

{We observed the nearby companion to TOI-5811 (here referred to as TOI-5811\,B) with the Calar Alto high-Resolution search for M-dwarfs with Exoearths with Near-infrared and optical \'Echelle Spectrographs (CARMENES) \citep{quirrenbach10} mounted at the 3.5\,m telescope of Calar Alto Observatory\footnote{{Program ID: DDT.26A.352 (PI: J.~Lillo-Box \& J.~Aceituno)}}. CARMENES is a fiber-fed \'echelle spectrograph with a visible and a near-infrared channel with resolutions of 94\,600 and 80\,400, respectively. For this work, we only used the data from the visible channel. Based on the detection of the nearby companion at 7.6 arcsec (reported in Sect.~\ref{sec:HRimaging}) and the contamination analysis presented in Sect.~\ref{sec:TESS}, showing potential for this companion to be the actual host of the transiting signals, we placed this nearby source (TOI-5811\,B) in fiber A of CARMENES. We illuminated fiber B with the Fabry-P\'erot to correct for the instrumental drift at the science frame compared to the wavelength calibration frame. We obtained a total of six spectra along the nights of 22-27 May 2026. The data were processed using the standard observatory pipeline \texttt{CARACAL} \citep{zechmeister14,bauer15} and the radial velocities were obtained using the \texttt{SHAQ} pipeline \citep{lillo-box22} through a cross-correlation of the extracted spectra with a G2 mask adapted from the ESPRESSO instrument pipeline \citep{pepe20}. The average uncertainty of the derived RVs is 4.2 m/s.}

%------------------------------
\subsubsection{NEID}
%------------------------------
We additionally observed TOI-4492 six times between 09-oct-2023 and 05-nov-2023 using the NEID spectrograph \citep{NEID_Schwab2016,NEID_Halverson2016}. NEID is a stabilized high-resolution \'echelle spectrograph on the WIYN 3.5\,m telescope at Kitt Peak National Observatory (KPNO) in Arizona (US). These observations were made with 120s exposure times in high-resolution (HR) mode ($R \sim 110\,000$). We reduced the NEID observations and extracted 1D spectra using v1.4.0 of the standard NEID Data Reduction Pipeline (DRP),\footnote{\url{https://neid.ipac.caltech.edu/docs/NEID-DRP/}}. As part of the NEID-DRP, RVs were derived {(with an average precision of 125 m/s)} with the cross-correlation method, using a weighted stellar line mask appropriate to the G2 spectral type of the star. {Because the large uncertainties, possibly due to telluric contamination, we re-derived these RVs by excluding specific orders that were noted to have telluric contamination and obtained an average error of 15 m/s.}

%------------------------------
\subsubsection{Tautenburg Echelle spectrograph (TLS)}
%------------------------------
We obtained 36 spectra of TOI-2114 using the Coud\'e \'Echelle spectrograph of the 2-m Alfred Jensch telescope of the Th\"uringer Landessternwarte (TLS) Tautenburg. Using the VIS-Gism, the spectra cover the wavelength-region from 452 to 765 nm. The resolution of the spectra is $R \sim 63\,000$ with the 0.52mm slit used. The observations were carried out using an iodine cell for wavelength calibrations. One spectrum was used as template, the others for RV-measurements with the iodine-cell. The exposure-times of the spectra taken with the cell were 1800s. The RVs were obtained from the reduced spectra using \texttt{viper} \citep{kohler25,viper}{, showing an average precision of 44 m/s.}

%++++++++++++++++++++++++++++++++++++++++++
\subsection{High-spatial resolution imaging}
%++++++++++++++++++++++++++++++++++++++++++
\label{sec:HRimaging}
High-angular resolution imaging is needed to search for nearby sources (typically down to $0.1\arcsec$) that can contaminate the TESS photometry, resulting in an underestimated planetary radius, or be the source of astrophysical false positives, such as background eclipsing binaries. In Table~\ref{tab:HRimaging}, we summarise the results of these observations, that are described along this section, and in Figs.~\ref{fig:tpfplots}-\ref{fig:tpfplots2} we display the sensitivity limits for each target and instrument used. 

%------------------------------
\subsubsection{`Alopeke at Gemini}
%------------------------------
Speckle interferometry of TOI-1137, and TOI-2114 was performed using the `Alopeke speckle instrument on the Gemini North 8-m telescope (\citealt{scott21}). `Alopeke provides simultaneous speckle imaging in two narrow bands (562 nm and 832 nm) with output data products including a reconstructed image with robust 5$\sigma$ magnitude contrast limits. 
All of the observations were reduced with our standard software pipeline described in \cite{howell11}.
TOI-1137 was observed twice on 12-oct-2019 and 13-sep-2022 with the same result showing TOI-1137 is a single star with no companions within the sensitivity limits (5-9 mag between $0.02\arcsec$ out to $1.2\arcsec$ at 5$\sigma$, corresponding to spatial limits of 8.4 to 503 AU).
TOI-2114 was observed on 26-may-2024 and was also found to be a single star with no companions within the sensitivity limits (5-8 mag between $0.02\arcsec$ out to $1.2\arcsec$ at 5$\sigma$, corresponding to spatial limits of 6.4 to 45.6 AU)

%------------------------------
\subsubsection{PHARO at Palomar and NIRC2 at Keck-II}
%------------------------------
Observations of TOI-603, TOI-1137, TOI-1837, TOI-2114, TOI-4492, TOI-5806, and TOI-5811 were made on 05-Dec-2020, 15-Sep-2025, 04-Jul-2025, 04-Jul-2025, 15-Sep-2025, 26-May-2024, and 08-Aug-2024, respectively, with the PHARO instrument \citep{hayward01} on the Palomar Hale (5m) behind the P3K natural guide star AO system \citep{dekany13}. Observations also were made of TOI-603 on 09-jun-2019 with the NIRC2 instrument on Keck-II (10m) behind the natural guide star AO system \citep{wizinowich00,schlieder21}.  The pixel scale for PHARO is $0.025\arcsec$ and for NIRC2 is $0.009942\arcsec$. The data were collected in a 9-point dither pattern in the $K_{cont}$ filter. The reduced science frames were combined into a single mosaiced image with final resolutions of $\sim 0.1\arcsec$ for PHARO and $0.05\arcsec$ for NIRC2.
The sensitivity of the final combined AO images were determined by injecting simulated sources azimuthally around the primary target every $20^\circ $ at separations of integer multiples of the central source's FWHM \citep{Furlan17}. The brightness of each injected source was scaled until standard aperture photometry detected it with $5\sigma $ significance. The final $5\sigma $ limit at each separation was determined from the average of all of the determined limits at that separation and the uncertainty on the limit was set by the rms dispersion of the azimuthal slices at a given radial distance.

%------------------------------
\subsubsection{ShARCS at Lick}
%------------------------------
Dressing et al. (submitted) observed TOI-1137, TOI-1837, and TOI-2114 on 15-oct-2019, 30-nov-2020, and 31-may-2021, respectively using the ShARCS camera on the Shane 3-meter telescope at Lick Observatory \citep{2012SPIE.8447E..3GK, 2014SPIE.9148E..05G, 2014SPIE.9148E..3AM}. Observations were taken with the Shane adaptive optics system in natural guide star mode in order to search for nearby, unresolved stellar companions. For all targets, they collected a single sequence of observations using a $Ks$ filter ($\lambda_0 = 2.150$ $\mu$m, $\Delta \lambda = 0.320$ $\mu$m). For TOI-1137 and TOI-2114, they also collected observations using a $J$ filter ($\lambda_0 = 1.238$ $\mu$m, $\Delta \lambda = 0.271$ $\mu$m). They reduced the data using the publicly available \texttt{SImMER}\footnote{\url{https://github.com/arjunsavel/SImMER}} pipeline \citep{2020AJ....160..287S}. They found no nearby stellar companions within their detection limits.

%------------------------------
\subsubsection{SAI}
%------------------------------
The objects TOI-1837, TOI-4492, and TOI-5806 were observed with the speckle polarimeter on the 2.5-m telescope at the Caucasian Observatory of Sternberg Astronomical Institute (SAI) of Lomonosov Moscow State University. 
For observations taken before 01-jul-2022 an electron multiplying CCD Andor iXon 897 was used as a detector \citep{Safonov2017}, later observations were conducted with a fast low-noise CMOS detector, the Hamamatsu ORCA -quest \citep{Strakhov2023}. The atmospheric dispersion compensator was active. The brighter targets were observed with a medium band filter centered at 625~nm with a 50~nm FWHM, while for the fainter ones, we used the standard $I_{\rm c}$ filter.For all observations, we estimated the detection limits from the autocorrelation function. For two targets, we detected a stellar companion: TOI-1837 and TOI-5806. The separation, position angle and magnitude difference were estimated by power spectrum approximation \citep{Strakhov2023}. The results are presented in Table~\ref{tab:HRimaging}. 
For TOI-1837, two epochs were obtained, separated by 0.56~yr, the expected shift with respect to the background is 28.5 mas based on Gaia DR3 proper motions, while the observed differential separation is much smaller. A similar situation is found for TOI-5806, with a timespan of 0.62~yr, and an expected shift of 20 mas, while no significant shift was detected. We then conclude that all two targets are binary stars.

%------------------------------
\subsubsection{NESSI}
%------------------------------
We observed TOI-1837 on 20-apr-2021 and TOI-4492 on 18-apr-2022 using the NN-EXPLORE Exoplanet Stellar Speckle Imager (NESSI; \citealt{scott18}) at the WIYN 3.5~m telescope on Kitt Peak. NESSI is a dual-channel speckle imaging system that delivers diffraction-limited resolution at optical wavelengths. For the TOI-1837 observation, we used filters centered on $\lambda_c = 562$~nm and $\lambda_c = 832$~nm. For the TOI-4492 observation, only the $\lambda_c = 832$~nm filter channel was available. Both observations consisted of a sequence of 9000 frames of 40~ms exposure per filter. NESSI's field-of-view was set by a $256\times256$ pixel sub-array readout, resulting in $4.6\arcsec\times4.6\arcsec$ images. However, we confined our contrast curve measurements to be within $1.2\arcsec$ from the target star to avoid the effects of decorrelation in speckle patterns at wider separations. Each observation was also accompanied by a set of $1000\times 40$~ms frames taken toward a nearby point source calibration star which served to measure the PSF at the time of the science observation.
In the case of TOI-4492, no companion stars were detected brighter than the contrast limits. A companion to TOI-1837 was detected at a position angle of 77$^{\circ}$, separation of $0.16\arcsec$ with ${\Delta}m=3.38$ at 562~nm and ${\Delta}m=2.79$ at 832~nm.

%------------------------------
\subsubsection{SOAR}
%------------------------------
% Carl Ziegler
We searched for stellar companions to TOI-5806 with speckle imaging on the 4.1-m Southern Astrophysical Research (SOAR) telescope \citep{tokovinin18} on 04-nov-2022, observing in Cousins I-band, a similar visible bandpass as TESS. This observation was sensitive to a 5.1-magnitude fainter star at an angular distance of $1\arcsec$ from the target. More details of the observations within the SOAR TESS survey are available in \cite{ziegler20}. An I-band 4.1 magnitude fainter companion was detected at a separation of $2.10\arcsec$ to TOI-5806 in the SOAR observations.

%++++++++++++++++++++++++++++++++++++++++++
\subsection{Ground-based photometry}
%++++++++++++++++++++++++++++++++++++++++++
Given the moderately large transit depths of the transiting candidates studied in this work, we only make use of the ground-based photometry, when available, to ensure that the transits occur on our target and to discard clear eclipsing binaries (EB) in nearby companions detectable through seeing-limited observations. A detailed description of these observations is provided in Appendix~\ref{app:GroundBasedPhotometry}.

%%===========================================================================
\section{Host star characterization}
%===========================================================================
\label{sec:host}

We determined the atmospheric parameters of the host stars ($T_{\rm eff}$, $\log g$, [Fe/H], $v\sin i$) via spectral synthesis, using the high-resolution spectra obtained with \cafe{} described in Sect.~\ref{sec:CAFEobs}. 
For each target, individual exposures were co-added to produce a single combined spectrum with typical signal-to-noise ratios of $\mathrm{S/N} \approx$ 100--280 per pixel in the 480--650\,nm range.

The analysis used \texttt{iSpec} \citep{Blanco-Cuaresma2014, Blanco-Cuaresma2019} with the \textsc{spectrum} radiative transfer code \citep{Gray1994}.
We leveraged the {\em Gaia}-ESO Survey (GES) atomic line list \citep[GESv6;][]{Heiter2021}, in combination with \textsc{marcs} model atmospheres \citep{Gustafsson2008} and the solar abundance scale of \cite{Grevesse2007}.
The synthesis was restricted to GES line regions at $R \sim 47\,000$. 
The H$\beta$, H$\alpha$, \ion{Na}{I}\,D, and \ion{Mg}{I}\,b lines were excluded due to non-LTE and chromospheric sensitivity, which are not captured by the 1D-LTE models employed by us. 
\ion{Ca}{II}\,H\&K and H$\gamma$ lie below the 440\,nm blue limit.

The macroturbulent velocity $v_{\rm mac}$ was held fixed at each iteration using the empirical calibration of \cite{Doyle2014}, valid for $5200 \leq T_{\rm eff}/\mathrm{K} \leq 6400$.
For hotter targets, the polynomial was extrapolated with the result clamped to $v_{\rm mac} \leq 8$\,km\,s$^{-1}$.
We derived $v\sin i$ both from spectral synthesis and from the CCF FWHM, adopting the latter as in the synthesis estimate cannot reliably separate $v\sin i$ and $v_{\rm mac}$.

Stellar luminosities were obtained by fitting spectral energy distributions (SEDs) to the BT-Settl CIFIST grid of theoretical spectra \citep{Baraffe1998, Allard2012} with VOSA \citep{Bayo2008}, following \cite{Cifuentes2020}.
The SEDs were derived from up to 15 photometric magnitudes (\textit{Gaia}~DR3, UCAC4, 2MASS, AllWISE) and parallactic distances from \textit{Gaia}~DR3.
Radii followed from the Stefan-Boltzmann law using $L_\star$ and the spectroscopic $T_{\rm eff}$.

For {isolated} main-sequence stars {(TOI-4492, and TOI-5806)}, masses were obtained from the empirical mass-luminosity relation (MLR) of \cite{Eker2018}, a piecewise power law calibrated on detached eclipsing binaries.
{Stellar radii were derived from the Stefan-Boltzmann law, and surface gravities were then computed from these masses and radii. Since $T_{\rm eff}$ enters both the spectral synthesis and the radius determination, the procedure was iterated until $\log g$ converged to within 0.01 dex. This approach breaks the age-$\log g$ degeneracy inherent to isochrone-based estimates for main-sequence stars, while still using the luminosity information through the empirical MLR and the radius determination.}

{For evolved targets with $\log g \lesssim 3.8$, the \cite{Eker2018} relation is no longer valid because luminosity rises at nearly constant stellar mass after the main-sequence turnoff. Likewise, for unresolved binaries, the luminosity is overestimated due to the contribution of the unresolved companion. Three targets (TOI-1137, TOI-1837, and TOI-5811\,B) fall into one or both of these categories. For these systems, we fitted $\log g$ as a free spectroscopic parameter through the Fe\,I/Fe\,II ionisation balance and adopted stellar masses from the empirical calibration of \citet{Torres2010}. This calibration is a polynomial in $T_{\rm eff}$, $\log g$, and [Fe/H], fitted to detached eclipsing binaries with dynamical masses, and is therefore independent of the potentially compromised luminosity estimates. Finally, TOI-2114 occupies an intermediate case: its luminosity is reliable, but its position in the HR diagram indicates that it is mildly evolved. For this target, we adopted the mass obtained by interpolating PARSEC v1.2S evolutionary tracks \citep{bressan12} in the $(T_{\rm eff}, L_\star)$ plane at the measured metallicity. For evolved or potentially blended systems, we also computed independent mass estimates from PARSEC isochrones and from the spectroscopic surface gravity combined with the Stefan--Boltzmann radius. These were not adopted as the final tabulated values, but used instead to verify that the reported masses were not biased by unreliable luminosity information or by the large uncertainty of spectroscopic $\log g$.}
The derived stellar parameters ($T_{\rm eff}$, $\log g$, [Fe/H], $v\sin i$, $L_{\star}$, $R_{\star}$, and $M_{\star}$) are listed in Table~\ref{tab:stellar_parameters}.

{Additionally, the} kinematic analysis with BANYAN~$\Sigma$ \citep{gagne18} classifies all seven targets as field stars ($P_{\rm field} > 99\,\%$), with no young association membership detected.

\begin{table*}[ht]
\caption{Derived stellar parameters. See Sect.~\ref{sec:host} for additional details.}
\label{tab:stellar_parameters}
\centering
\small
\renewcommand{\arraystretch}{1.12}
\begin{tabular}{l c c c c c c c c}
\toprule
Target & SpT$^{\dagger}$ & {$T_{\rm eff}$} & {$\log g$} & {[Fe/H]} & {$v\sin i$}$^{\ddagger}$ & {$L_{\star}$} & {$M_{\star}$} & {$R_{\star}$} \\
 & & \multicolumn{1}{c}{(K)} & \multicolumn{1}{c}{(dex)} & \multicolumn{1}{c}{(dex)} & \multicolumn{1}{c}{(km s$^{-1}$)} & \multicolumn{1}{c}{($L_\odot$)} & \multicolumn{1}{c}{($M_\odot$)} & \multicolumn{1}{c}{($R_\odot$)} \\
\midrule
TOI-603 & F8IV & 6027 $\pm$ 93  & 4.255 $\pm$ 0.042 & 0.183 $\pm$ 0.055  & 4.196 $\pm$ 0.069 & 2.592 $\pm$ 0.023 & 1.240 $\pm$ 0.092 & 1.477 $\pm$ 0.046 \\
TOI-1137 & K0III+ & 5041 $\pm$ 63  & 3.24 $\pm$ 0.20 & --0.350 $\pm$ 0.052 & $< 3$ & 20.83 $\pm$ 0.23$^c$ & 1.43 $\pm$ 0.16 & 5.98 $\pm$ 0.15 \\
TOI-1837 & F6V+ & 6340 $\pm$ 140 & 4.46 $\pm$ 0.20 & 0.168 $\pm$ 0.066  & 11.187 $\pm$ 0.045 & 3.90 $\pm$ 0.17$^a$ & 1.24 $\pm$ 0.10$^{b}$ & 1.08 $\pm$ 0.27$^{b}$ \\
TOI-2114 & F5IV & 6400 $\pm$ 170 & 4.03 $\pm$ 0.17 & 0.001 $\pm$ 0.078  & 10.728 $\pm$ 0.047 & 5.894 $\pm$ 0.074 & 1.309 $\pm$ 0.052 & 1.97 $\pm$ 0.11 \\
TOI-4492 & G2V & 5711 $\pm$ 83  & 4.363 $\pm$ 0.034 & 0.219 $\pm$ 0.070  & $< 3$ & 1.4304 $\pm$ 0.0082 & 1.067 $\pm$ 0.055 & 1.222 $\pm$ 0.036 \\
TOI-5806 & F3V & 6730 $\pm$ 190 & 4.256 $\pm$ 0.059 & --0.019 $\pm$ 0.099  & 19.072 $\pm$ 0.042 & 3.547 $\pm$ 0.022 & 1.333 $\pm$ 0.099 & 1.387 $\pm$ 0.080 \\
TOI-5811A & G0IV+ & 6090 $\pm$ 140 & 3.74 $\pm$ 0.21 & --0.049 $\pm$ 0.086 & 8.750 $\pm$ 0.046 & 11.14 $\pm$ 0.26$^a$ & 1.46 $\pm$ 0.212 & 2.57 $\pm$ 0.93 \\ 
TOI-5811B & K2IV+ & 4840 $\pm$ 110 & 3.33 $\pm$ 0.12 & --0.06 $\pm$ 0.10 & $< 3$ & 6.877 $\pm$ 0.062 & 1.08 $\pm$ 0.32 & 3.73 $\pm$ 0.17 \\ 
\bottomrule
\end{tabular}
\tablefoot{
$^{(\dagger)}$ Plus sign indicates known unresolved multiplicity.
$^{(\ddagger)}$ For slow rotators ($v\sin i < 5 $~km/s), this parameter is actually a mixture of the rotational broadening, the microturbulence velocity and the macroturbulence contributions.
$^{(a)}$ Luminosity overestimated due to unresolved companion. 
$^{(b)}$ {For TOI-1837, the adopted Torres et al. (2010) mass and radius depend on the spectroscopic $\log g$, which may be marginally overestimated because of rotational line broadening. The quoted uncertainties account for this effect.}
$^{(c)}$ Despite the presence of an eclipsing binary companion ($P = 1.35$\,d, $\sim$2000\,ppm), the luminosity is unaffected as the companion contributes $\lesssim$1\% of the total optical flux of the K0 giant, consistent with the observed eclipse depth.
}
\end{table*}

%===========================================================================
\section{Methodology}
%===========================================================================
\label{sec:analysis}

%++++++++++++++++++++++++++++++++++++++++++
\subsection{Radial velocity and light curve model definition}
%++++++++++++++++++++++++++++++++++++++++++
In this section we perform a joint analysis including the TESS light curve and the radial velocities gathered. In general terms, we model both datasets assuming the presence of one planetary signal with the period of the reported TESS signal. In order to perform this analysis, we use our own code \texttt{rebeal}, previously used in other works (e.g., \citealt{lillo-box20a}).

\rebeal{} is a wrap up Python package allowing the user to model either independently or jointly radial velocities, transit signals and phase curves (including the reflexion, ellipsoidal and beaming effects; REBs, see \citealt{lillo-box14}). It uses the \texttt{batman} algorithm \citep{batman} to model the transit signal, which for the uniform, linear and quadratic limb darkening formulations uses the \cite{mandel02} formalism to calculate the transit shape. \rebeal{} allows for different sets of parametrization, depending on the inclusion or not of the REB effects. These have dependencies with the planetary mass and radius different than for modeling transits and RV. Hence, if REBs are included in the model, then \rebeal{} parametrizes the problem by explicitly including the planetary and stellar mass and radius as independent parameters in the modeling. Otherwise, \texttt{rebeal} also allows for a relative parametrization consisting on the orbital period ($P$), time of conjunction (T$_0$), planet-to-star radius ratio ($R_p/R_{\star}$), RV semi-amplitude ($K$), stellar density ($\rho_{\star}$), orbital inclination ($i$), and a quadratic limb darkening formulation ($u_1$ and $u_2$). For eccentric cases, we also include the eccentricity ($e$) and argument of the periastron ($\omega$). Added to this, we include a planetary geometric albedo ($A_g$) to test the possibility of secondary eclipses. Added to these parameters, we include a photometric jitter ($\sigma_{\rm TESS}$) and offset ($\delta_{\rm TESS}$), and a radial velocity jitter and offset per instrument. We also add a slope parameter to account for possible long-term RV trends. 

The \rebeal{} code  also allows for the simultaneous modeling of the out-of-transit phases using Gaussian Processes (GPs) under different possible kernels included in either \texttt{george} \citep{george}, \texttt{celerite} \citep{celerite1} or \texttt{celerite2} \citep{celerite2} implementations. For the cases presented in this paper, we use a \texttt{Matern-3/2} kernel from \texttt{celerite2} to model the additional flux variations due to either residual instrumental systematics or stellar variability. This kernel is sufficiently flexible to model short term variability and depends on two hyper-parameters, namely the amplitude of the GP ($\eta_{\sigma}$) and the time scale ($\eta_{\tau}$). However, in the case of TOI-4492, the clear stellar pulsations made it necessary to use a \texttt{SHOTerm} kernel to better model the high-frequency flux variability. A detailed study of these oscillations is out of the scope of this work. 

{For the radial velocities, stellar activity is not a critical point in these systems because i) except for TOI-5806, there is no evidence of a radial velocity correlation with the available activity indicators {(in all cases showing Pearson's r correlation coefficient values below 0.4, indicating no or very weak correlation with the FWHM for CAFE, CARMENES and HARPS data, and BIS for SOPHIE)}, ii) the RV amplitude of the planet candidates are large enough to not be affected by stellar activity unless a very clear correlation is present (which is not the case in our datasets), and iii) the transiting nature of the candidates already provides an independent measurement of the orbital period, {thereby minimizing potential confusion between the planetary signal and the stellar rotation period}. Consequently, all systems analysed in this work (except TOI-5806) do not require an explicit analysis of the stellar activity.} For TOI-5806 we required a GP to jointly model the RVs and the BIS activity indicator from the SOPHIE data {(see Sect.~\ref{sec:SOPHIE})}. In this case, we use a quasi-periodic kernel as described in \cite{faria16} and its implementation in the \texttt{george} package \citep{george}.

%++++++++++++++++++++++++++++++++++++++++++
\subsection{Parameter inference and Bayesian evidence}
%++++++++++++++++++++++++++++++++++++++++++

Based on the above parametrization, we sample the parameter space and populate the posterior distribution by using the implementation of the \cite{goodman10} affine invariant ensamble sampler python module \texttt{emcee} \citep{emcee}. Because these are transit candidates, we use the Gaussian priors on the periods and times of conjunction centered on the transit values but use extended amplitudes for their distributions corresponding to 10 times the reported uncertainties. We note that when adopting more uninformative priors, the results are nearly unchanged. Since here we are not aiming at probing the statistical significance of the signals (which is large enough given the large transit depths), the simplification of assuming Gaussian priors does not affect the final results while it speeds up the convergence of the method. We use four times the number of free parameters as the number of walkers for the MCMC sampling, with each walker running for $10^4$ steps in a first burn-in phase. A second run is then initialized in a ball around the maximum a posteriori region of the parameter space obtained from the burn-in phase. All samples from this latter phase are then used to build up the marginalized posterior distributions from where we obtain the median (50\% percentiles) and 68.7\% confidence interval for each parameter. The convergence of the chains is ensured by imposing that the length of the chain is larger than 50 times the autocorrelation time of the chain.

We performed model comparison among different scenarios, including different planetary architectures with circular or eccentric orbits and different number of Keplerians in the system, through the estimation of the Bayesian log-evidence ($\Delta\ln{\mathcal{Z}}$) using the \texttt{bayev}\footnote{\url{https://github.com/exord/bayev/tree/master}} code (\citealt{bayev}) based on the importance sampling estimator introduced by \cite{perrakis14}. We consider a strong preference for a more complex model against a simpler one if $\Delta\ln{\mathcal{Z}} > 6$ based on the classification from \cite{jeffreys98}.
 
%++++++++++++++++++++++++++++++++++++++++++
\subsection{Assessment of the planetary nature}
%++++++++++++++++++++++++++++++++++++++++++
\label{sec:GenericPrinciples}

There are ongoing discussions in the exoplanet community about developing an Exoplanet Confirmation Protocol (ECP, Lillo-Box et al., in prep.; already applied in planet-discovery papers, see \citealt{balsalobre-ruza25}) that would define the minimum requirements for a detected signal to qualify as a ``confirmed exoplanet.'' In the proposed scheme, three generic principles (GP1, GP2, and GP3) must be demonstrated for a signal to be considered as a bona fide confirmed exoplanet: 
\begin{enumerate}
	\item Signal and model significance (GP1): The signal is statistically significant, with the relevant properties of the planet (e.g., radius and mass) being identified at a statistically significant level (in this paper defined as $3\sigma$). Also, the model of planet hypothesis should be statistically preferred against the null hypothesis (i.e., no planet\footnote{{Note that the null hypothesis here refers to a model with no Keplerians but including, if relevant, the stellar activity model.}}), for instance through model comparison using the Bayesian evidence or other justified metrics. In this paper we use the Bayesian evidence as a metric to demonstrate this point. \vspace{1pt}
	\item Demonstrated origin of the signal (GP2): The star from which the signal originates (i.e., the host star) is known confidently. In this paper, we use both high-resolution images, centroid motion analysis on the TESS data, ground-based seeing limited observations, and spectroscopic arguments to pinpoint the origin of the transiting signal. \vspace{1pt}
	\item Planetary-mass regime (GP3): Confirming that the signal comes from an object in the planetary-mass domain to ensure it conforms with the IAU working definition of extrasolar planets (see \citealt{lecavalier22}); in particular, the mass must be compatible at a certain degree of confidence with being smaller than 13~\Mjup.
\end{enumerate}

In the next section, we present (based on radial velocities and transit data) the evidence for or against each of the three criteria described above, for each target.

%++++++++++++++++++++++++++++++++++++++++++
\section{Results}
%++++++++++++++++++++++++++++++++++++++++++
\label{sec:results}

\begin{figure*}
	\includegraphics[width=0.95\textwidth]{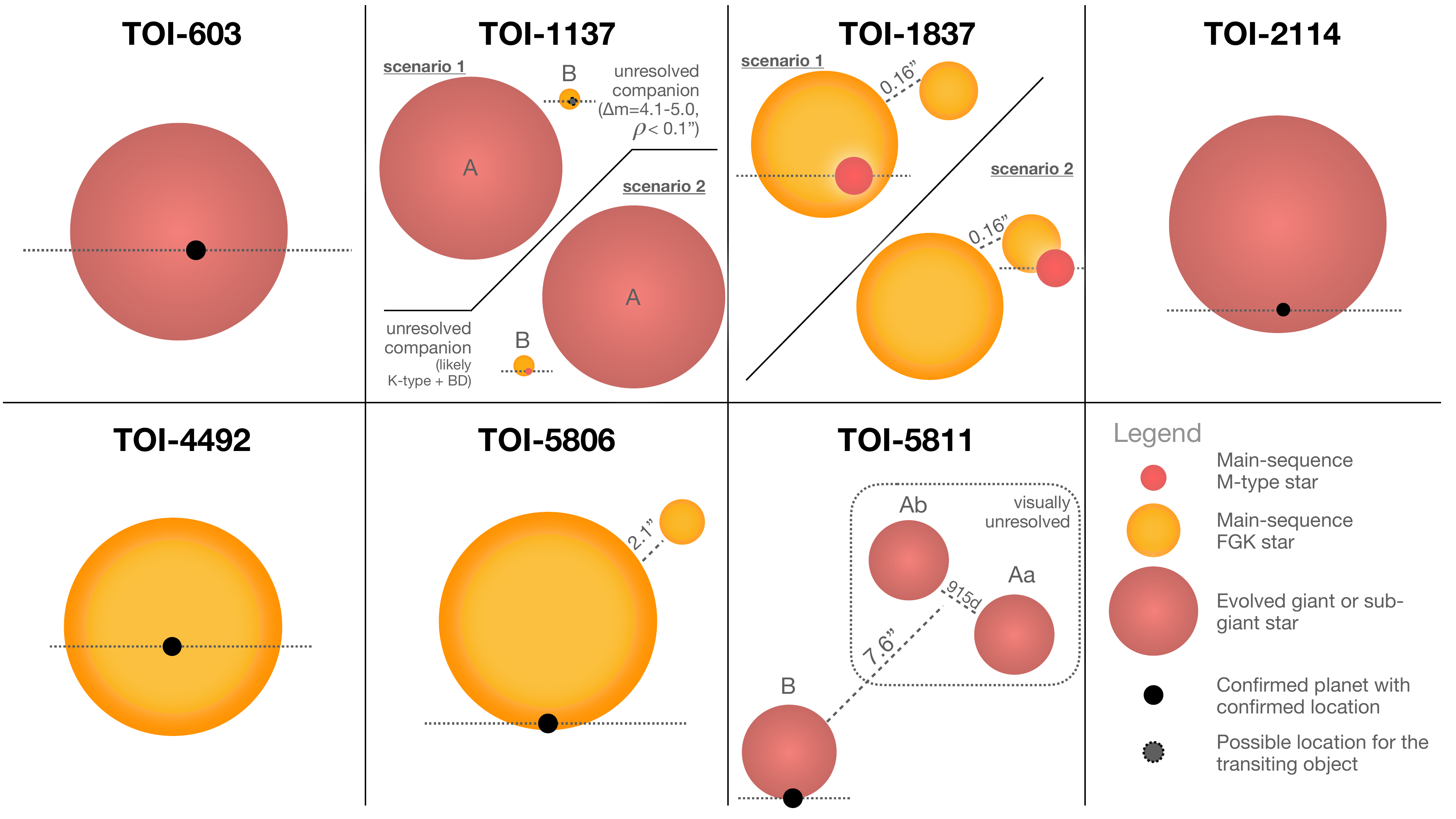}
	\caption{Schematic view of the resulting configurations inferred by our study of the seven transiting candidates. Confirmed planets with confirmed locations are shown as black circles with sizes proportional to their host star symbol size. The location of the planet symbols also corresponds to the inferred impact parameter. Dotted lines indicate the motion across the transit. }
	\label{fig:schematic}
\end{figure*}

These results show a clear detection of the RV signal of {five} planets among the seven studied in this paper. In particular, TOI-603.01, TOI-2114.01, TOI-4492.01, TOI-5806.01, {and TOI-5811B.01} show statistically significant radial velocity variations in phase with the transit ephemeris. In Table~\ref{tab:posteriors}, we present the priors used for each parameter and the final median and confidence intervals obtained from the marginalised posterior distributions. The confidence on the planetary confirmation of each candidate, following the generic principles described in Sect.~\ref{sec:GenericPrinciples}, is discussed individually in the following sections. In Fig.~\ref{fig:schematic}, we present a schematic representation of the configurations for each of the seven targets discussed in this paper after the analysis of the data in hand in order to guide the reader through the discussion. 

%------------------------------
\subsection{TOI-603}
%------------------------------

This target was initially included in our \cafe{} program because its originally reported orbital period of 8~d for the transiting signal fitted into the constraints of our survey. Indeed, we obtained 15 \cafe{} RVs between nov-2021 and feb-2022. However, after the release of sector 46 (jan-2022), the period determination was revised to be 16~d. With the revised period, the CAFE$_2$ precision is insufficient to constrain the planet signal and so we do not use these data in our analysis. Instead, we use the RVs from HIRES presented in \citep{chontos24}\footnote{Since at the time of writing the paper by \cite{chontos24} is still in review process, the data was directly provided by the authors of such publication and is still not publicly available} and HARPS (PI: Louise Nielsen). The RVs from these two datasets are consistent with the period and phase implied by the transit ephemeris, thus suggesting no need for the inclusion of GP modeling of the RV to confirm the candidate. We then proceed with the joint analysis and find that the transiting signal is explained by a sub-Saturn-like planet with a mass of $m_b=35.7^{+6.5}_{-6.5}$~\Mearth{} and a planet radius of $R_b=7.46^{+0.19}_{-0.21}$~\Rearth. The priors and posterior distributions for the parameters as well as the derived properties inferred from them are shown in Table~\ref{tab:posteriors}. The phase-folded RVs and transit data and model are shown in the left panel of Fig.~\ref{fig:result-TOI2114}.  \\

%------------------------------
\subsection{TOI-1137}
%------------------------------
\label{sec:TOI-1137}

The spectroscopic analysis of TOI-1137 (see Sect.~\ref{sec:host}) implies an evolved sub-giant or RGB star, with a radius of  $5.98\pm0.15$~\Rsun{} (in agreement with the radius reported in \gaia{} DR3). This is incompatible with the observed $\sim$1.6 hours {transit duration} (unless in case of a grazing eclipse) and, more importantly, with the short orbital period implying an orbital radius of $\approx$5.5~$\Rsun$, i.e., likely inside the star.  Consequently, the transiting object cannot orbit the evolved star (hereafter referred to as star A) and there must be another star (B) contributing to the TESS photometric signal, likely being the source of the transits. Furthermore, the TESS light curve shows clear secondary eclipses of depth of $\sim$300~ppm, as well as out-of-transit phase-curve variations of amplitude $\sim 40$~ppm (see Fig.~\ref{fig:TOI1137_EB}). Hence, the nature of the transiting object (component C) is uncertain. 
Added to this, 
i) the CCFs do not show evidence for a secondary peak, 
ii) there are no additional sources detected neither in the high-spatial resolution images within the sensitivity limits {nor in the \cite{Gaia} RUWE value, reported to be 0.923 (consistent with no additional blended sources)}, and 
iii) the ground-based (seeing-limited) observations ruled out transits of the evolved star (see Sect.~\ref{app:GroundBasedPhotometry}). 
All these findings point to a very close, relatively faint, photometrically and spectroscopically unresolved companion (B) as the source of the transits. The next question is whether the transiting companion of star B is a planet, or not. In Appendix~\ref{app:TOI-1137} we describe the evidences we have for the different possible scenarios with component C being a planet or a low-mass star or brown dwarf  around component B. Yet, with the data in hand, we cannot provide a definitive answer.

%------------------------------
\subsection{TOI-1837}
%------------------------------

The \cafe{} radial velocities of this target are in phase with the TESS transit ephemeris. The standalone RV analysis shows a preference ($\Delta\ln{\mathcal{Z}}=+38$) for a slightly eccentric orbit ($e=0.0344\pm 0.0050$), with a RV semi-amplitude of 18.5 km/s. {This would correspond to a minimum mass for the companion (assuming a host mass as derived in Sect.~\ref{sec:host}) of 0.18~\Msun, thus in the stellar regime}. The CCF shows a moderately broadened profile, with an average FWHM of 17.6 km/s, which is partly correlated with the RVs. This may indicate the presence a second component in the CCF, {which is also hinted by the large reported \textit{Gaia} RUWE of 7.37}. The companion star at $0.16\arcsec$ separation (see Sect.~\ref{sec:host}) could be the responsible for this correlation. Given that the transit-like signal is definitely not due to a planet, we elected not to pursue further analysis.

%------------------------------
\subsection{TOI-2114}
%------------------------------
For TOI-2114, we first tested three RV-only models, including no planets ($0p$), one planet in circular orbit ($1p1c$), and one planet in eccentric orbit ($1p$). 
It is important to note at this point that the number and precision of radial velocity datapoints is insufficient to obtain an independent detection of the planet via RVs. Still, the absence of companions within our sensitivity limits allows us to associate the transits with the target star and hence the inclusion of Gaussian priors in the RV-only analysis is justified. The results of this analysis show a strong preference of the $1p$ model against the other two, with Bayesian evidence differences of $\ln{\mathcal{Z}_{1p}}-\ln{\mathcal{Z}_{0p}} = + 16.3$; and also a clear preference of the eccentric versus the circular model, with $\ln{\mathcal{Z}_{1p}}-\ln{\mathcal{Z}_{1p1c}} = + 17.9$. 

Subsequently, we perform the joint modeling using both the TESS light curves and the RV data, now using broad uninformative priors on the period and ephemeris. Here we only test the $1p$ model, as the evidence of the presence of the transit is clear. Table~\ref{tab:posteriors} shows the priors and posterior values for this analysis, as well as derived parameters using also the stellar properties derived in Sect.~\ref{sec:host}. As shown, we can identify the nature of the transiting object as a giant planet (TOI-2114\,b) with  $R_p=1.271^{+0.059}_{-0.059}$~\Rjup{} and $m_p= 1.01^{+0.14}_{-0.12}$~\Mjup, with an orbital semi-major axis $a=0.0752^{+0.0046}_{-0.0043}$~au ($a/R_{\star} = 7.71^{+0.32}_{-0.28}$) with an eccentricity of $e=0.487^{+0.028}_{-0.030}$. Figure~\ref{fig:result-TOI2114} (middle panels) shows the data and median model. 

The mass and radius of the transiting object are measured with 7.2$\sigma$ and 208$\sigma$ confidence. Also, as shown above, the planet model (even in the RV standalone analysis) provides a Bayesian evidence larger than our threshold ($\Delta\ln{\mathcal{Z}}>+6$) against the null hypothesis. Both metrics ensure the accomplishment of GP1. The absence of evidence for other sources given by the combination of high-resolution images, a \textit{Gaia} RUWE metric of 1.108, and a null correlation between the RV and the FWHM indicator ensures that the origin of the signal is the target star, thus accomplishing GP2. Finally, the inferred  mass of the transiting object, $m_p= 1.01^{+0.14}_{-0.12}$~\Mjup, satisfies GP3. Consequently, we can confirm the planetary nature of this object, that will hereafter be named TOI-2114\,b.

%------------------------------
\subsection{TOI-4492}
\begin{figure}
\centering
	\includegraphics[width=0.48\textwidth]{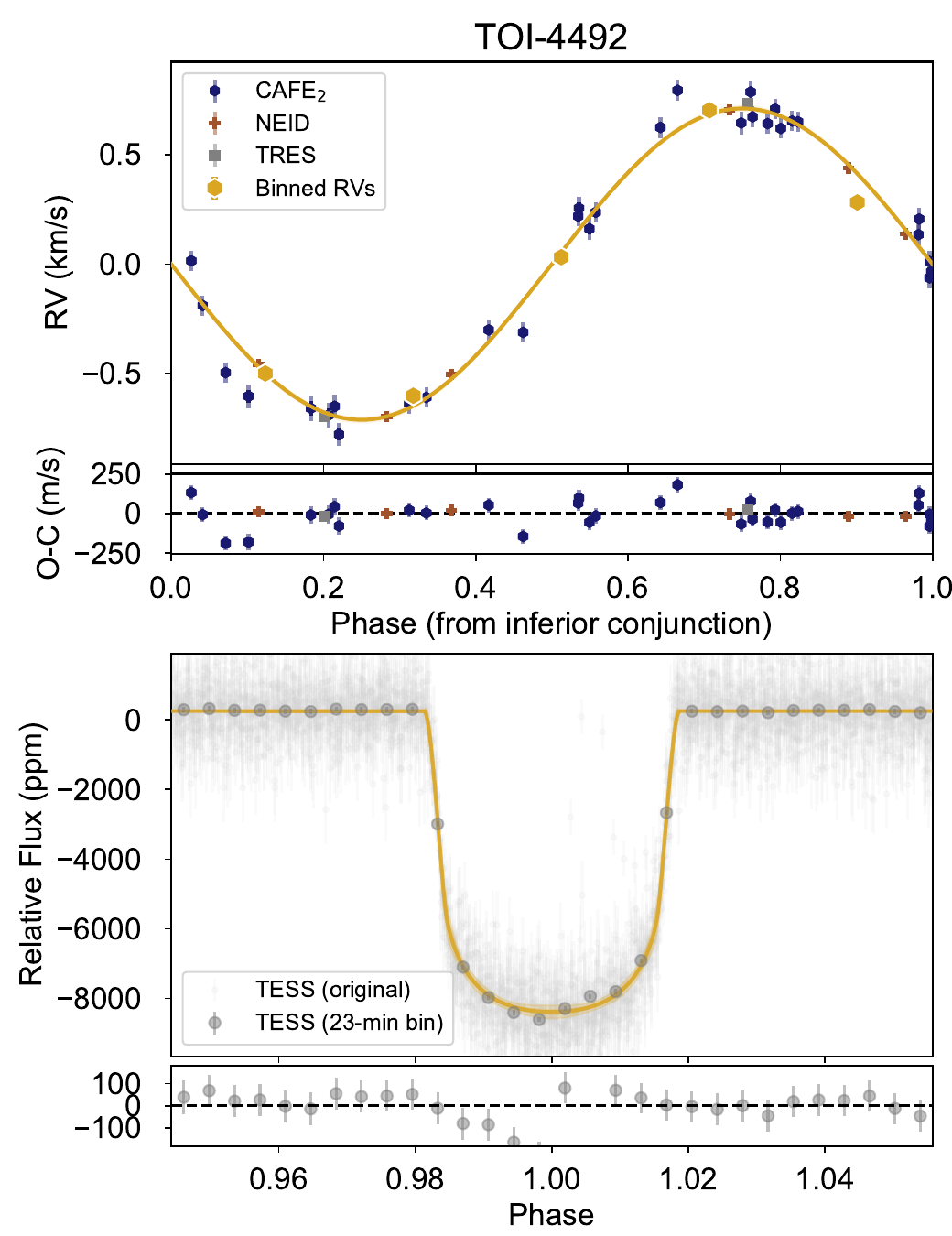}
	\caption{Results for TOI-4492\,b from the joint radial velocity and light curve analysis. \textbf{Upper panels:} Phase-folded radial velocity observations including all instruments used in the modeling (see legend). RVs binned into five orbital phases with a weighted average method are shown for reference as gold hexagons. The median RV model is shown as a solid gold line and the 68.7\% confidence interval is represented by the shaded gold region. \textbf{Lower panels:} Zoom-in to the transit region. Original cadence TESS data is shown as light gray small symbols, while dark gray circles represent 14-min bins in phase. The final median model is shown as a solid gold line and the 68.7\% confidence interval is represented by the shaded gold region.}
	\label{fig:result-TOI4492}
\end{figure}

%------------------------------
For TOI-4492, we also tested the $0p$, $1p1c$ and $1p$ hypotheses. In this case, again, the star is isolated so we can use the priors on the period and ephemeris from the transit signal. Yet, we tested the use of uninformative priors and the Bayesian evidence is strongly significant in both cases in favor of the planet model. In particular, in this case, the most favored solution corresponds to the circular case, with $\ln{\mathcal{Z}_{1p1c}}-\ln{\mathcal{Z}_{0p}} > 1000$. We then proceed with the joint LC and RV modeling to obtain the final parameters presented in Table~\ref{tab:posteriors}. In summary, the transiting object (TOI-4492\,b) corresponds to a Jupiter-size planet with $R_p=1.051 \pm 0.055$~\Rjup{} and a mass of $m_p=5.92^{+0.67}_{-0.64}$~\Mjup{} in a circular orbit with a tight separation of $a/R_{\star}=9.17^{+0.15}_{-0.25}$. Figure~\ref{fig:result-TOI4492} shows the phase-folded RV and transit curve with the corresponding models.

Similarly to the previous target, in this case we also satisfy GP1 with 19$\sigma$ and 8.8$\sigma$ significance on the radius and mass (respectively) of the transiting object and a Bayesian evidence of the planet model of $\Delta\ln{\mathcal{Z}}>+6$ against the null hypothesis. GP2 is also accomplished through similar terms (based on the absence of detected companions capable of mimicking the transiting signal, and no astrometric variability {with a \textit{Gaia} RUWE of 1}). Finally, GP3 is also accomplished with a mass of $m_p=5.92^{+0.67}_{-0.64}$~\Mjup{}. Hence, we confirm the planetary nature of this object and will refer to it as TOI-4492\,b throughout this paper. 

%------------------------------
\subsection{TOI-5806}
\label{sec:TOI-5806}
%------------------------------
TOI-5806 has a companion (\gaia{} DR3 1780750207504665600) at separation $2.1\arcsec$ and magnitude difference $\Delta m=5.7$~mag at 635nm (see Sect.~\ref{sec:observations}). It has a proper motion and parallax compatible with those of the target star, and hence is very likely a gravitationally bound companion. The ground-based, seeing-limited observations presented in Sect.~\ref{app:GroundBasedPhotometry} are able to separate this from the main target and confirm that the transits occur on the main target (TOI-5806). The second star is too faint ($\Delta G = 3.3$~mag, $\Delta I_c = 4.1$~mag) compared to the primary star to produce a significant RV shift. Also, the on-sky diameter of the \cafe{} fiber corresponds to $2.4\arcsec$, hence, the companion is $\sim 0.9\arcsec$ away from the fiber border and should produce a negligible contribution to the spectrum with the main target centered on the fiber. A similar argument can be made for SOPHIE observations (with the SOPHIE on-sky aperture being $3\arcsec$). Indeed, {no correlation between the FWHM and the RVs is found} with a Pearson's correlation coefficient of 0.03 and a corresponding p-value of 0.84. Hence, any RV variations are in this work attributed to the main target. 

The larger time span of the SOPHIE observations (519~d) versus the \cafe{} data (359~d) allows us to identify a long period signal in the periodogram, in the range 400-800 days. However, this long-periodicity is also visible in the BIS time series. Additionally, a shallow 14~d periodicity is also visible in the BIS time series that could be attributed to the stellar rotation period. We thus proceed with the standalone analysis of the RVs and the BIS data but in this case adding a GP with a quasi-periodic kernel. We leave the amplitude of the kernel ($\eta_1$) with an uninformative prior (one for the RVs and one for the BIS) and the characteristic rotation period ($\eta_3$), time scale of the variations ($\eta_2$) and weight among both time scales of the kernel ($\eta_4$) being shared by both the RV and BIS time series, also with uninformative priors. 

In this case, we test three models, including 0-planet, 1-planet (the transiting planet) and 2-planets (with a second, possibly fictitious planet to account for the long-period signal). The one-planet model is clearly favored against the null hypothesis ($\ln{\mathcal{Z}_{1p1c}}-\ln{\mathcal{Z}_{0p}} = +16$) and, in particular, the circular model is preferred against the eccentric model by $\ln{\mathcal{Z}_{1p}}-\ln{\mathcal{Z}_{1p1c}} = -1.6$, thus allowing us to set an upper limit of $e<0.1$ at 95\% confidence. Moreover, the current data are not able to support the two-planet hypothesis for the long-period signal, which is here explained by the long-term BIS variability (likely related to the magnetic cycle). Still, since the evidence of the two-planet model is at odds with the one-planet hypothesis, it is worth reporting that such a second potential planetary signal converges to a periodicity of $523.6^{+4.8}_{-2.6}$~d in an eccentric ($e=0.56\pm0.07$) orbit. However, since the Bayesian evidence is not statistically supporting the second planet hypothesis, {we only assume the presence of one Keplerian in the joint analysis}. It is also worth reporting that the rotation period converges to $14.217\pm0.087$~day, showing the relatively rapid rotation of the star, compatible with the observed rotational broadening of the CCF (see Sect.~\ref{sec:host}). 

 We then proceed to the joint analysis under the one planet circular orbit scenario but including the quasi-periodic GP for the RVs and BIS. Figure~\ref{fig:result-TOI2114} (right panels) shows the phase-folded RV and transit curve with the corresponding models and Table~\ref{tab:posteriors} shows the full posterior (and prior) values for this modeling. 
 In summary, the data are compatible with a $R_p=1.35^{+0.17}_{-0.11}$~\Rjup{} planet with a mass of  $m_p=2.27^{+0.26}_{-0.25}$~\Mjup{} in a nearly circular orbit. The expected amplitude of the ellipsoidal variations for such a planet is $8.9\pm 4.2$~ppm, hence undetectable for TESS.

In this case, as in the previous two cases, the significance of the signal and model satisfy the criterion of GP1. Despite the presence of a close companion, we have provided sufficient evidence based on ancillary observations (including high-resolution imaging, centroid motion, and probabilistic arguments described in Sect.~\ref{sec:TESS}) that the transits occur on the main (bright) target and that the RV variations at the transit period are produced by the main target (minimal contamination on the fiber as the companion lies far from the fiber aperture). Hence, GP2 is also ensured in this case. Finally, GP3 is demonstrated based on the mass of $m_p=2.77^{+0.34}_{-0.32}$~\Mjup{} obtained from the joint analysis. As a consequence, the transiting planet candidate is shown to be a genuine planet that we hereafter name TOI-5806\,b\footnote{This planet was previously detected by the SuperWASP survey. However, a description of this object has not heretofore been published because the team did not have quite enough data to formally confirm the candidate (Coel Hellier, private communication).}.

%------------------------------
\subsection{TOI-5811}
%------------------------------
\label{sec:results-5811}

TOI-5811 is a hierarchical triple star, with a spectroscopic binary unresolved by \gaia{} (SB1{, with a RUWE of 1.84}) but spectroscopically resolved by our \cafe{} observations (SB2), with a period of 916\,d (components {Aa and Ab}), plus an external third component ({B}; TIC\,331484427) located at 7.6\arcsec{} and with proper motions compatible with those of components Aa and Ab (as shown by \gaia{} and displayed in the corresponding panel of Fig.~\ref{fig:tpfplots}). With \cafe, we observed the components of the SB1 and we were capable of separating the RVs of both stars. We modeled the RVs of components {Aa and Ab} both assuming only the presence of the other star (models named 1pAa and 1pAb respectively for each component) and also including the transiting planet candidate's period and reference epoch (models named 2p2cAa and 2p2cAb) with Gaussian priors. In all cases, we use the period and eccentricity values from the \gaia{} NSS catalog and their uncertainties as Gaussian priors in the analysis. The results show a clear preference for the models not including the planetary signal in any of the two RV datasets (from stars Aa and Ab). 

{With the centroid motion analysis in Sect.~\ref{sec:TESS} providing a probability of 35.9\% for the third component (TOI-5811\,B) being the host of the transits and the ground-based photometry data reported in Sect.~\ref{app:GroundBasedPhotometry} also suggesting this possibility (although the lack of pre-ingress data complicates the interpretation of the results), we pointed the CARMENES (6 spectra along 6 consecutive nights) and TRES (2 spectra) instruments to this alternative host star. Given the relatively large separation (7.6 arcsec) from the components of the spectroscopic binary (Aa and Ab), the contamination from these two sources in the fiber is negligible. The RVs from both instruments are in-phase with the transit ephemeris, thus showing that this is actually the source of the transiting signal. We use these RVs and the TESS lightcurve to perform a new analysis and obtain an inferred planetary mass and radius of $0.81^{+0.11}_{-0.10}$~\Mjup{} and $1.96^{+0.18}_{-0.18}$~\Rjup, respectively.}

{With this analysis, we can now confirm the planetary nature of the transiting object but around TOI-5811\,B, since we have demonstrated 
i) GP1 through the significance of the model against the null hypothesis ($\ln{\mathcal{Z}_{1p1c}}-\ln{\mathcal{Z}_{0}} > +400$) and the statistically significant mass and radius detection,
ii) GP2 through the observation of the alternative source of the transits and the consequent recovery of the RV signal compatible with the transit ephemeris, and
iii) GP3 through a mass determination in the planetary regime ($m_p\sim0.81$~\Mjup{}). 
}

%________________________________________________________________
\section{Discussion}
\label{sec:discussion}

\begin{figure*}
	\includegraphics[width=1\textwidth]{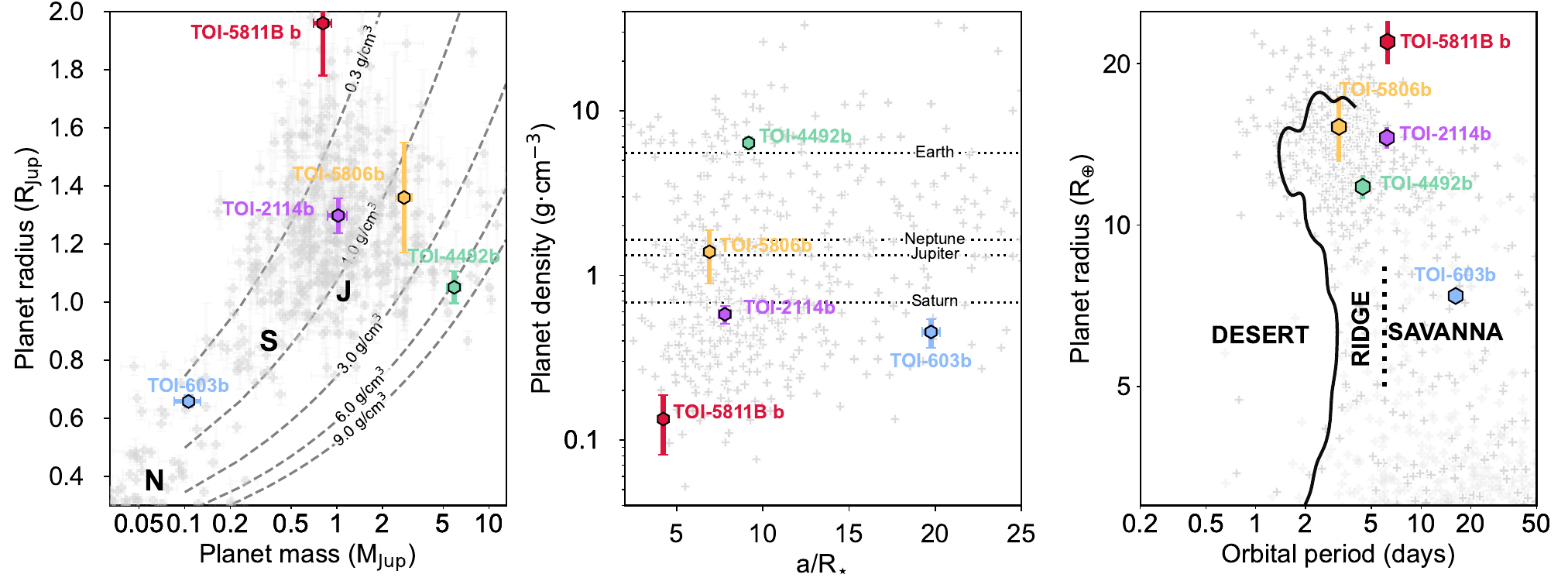}
	\caption{Confirmed planets from this study in context with the current exoplanet population. \textbf{Left panel}: Mass-radius diagram for known planets (gray plus symbols) and our newly confirmed planets (colored hexagons). Iso-density lines are included and labelled. The location of Jupiter (J), Saturn (S) and Neptune (N) are also indicated. \textbf{Middle panel}: Planet density versus star-planet separation ($a/R_{\star}$). TOI-603\,b stands out as a puffy planet, unusual for its relatively large separation (see Sect.~\ref{sec:discussion} for discussion). \textbf{Right panel}: Planet radius versus orbital period for the sample of known planets. Planets with well derived ($>3\sigma$) masses are shown with darker gray color. The location of the dessert, ridge and savanna as defined in \cite{castro-gonzalez24} are marked (we used the \texttt{nep-des} code - \url{https://github.com/castro-gzlz/nep-des}- to build this plot).}
	\label{fig:mass-radius}
\end{figure*}

{Among the five confirmed planets, four} (TOI-2114\,b, TOI-4492\,b, TOI-5806\,b, {and TOI-5811\,B\,b}) are hot Jupiters with well determined masses and radii, while TOI-603\,b is a super-Neptune planet in the savanna \citep{bourrier23}. The left panel in Fig.~\ref{fig:mass-radius} shows the mass-radius diagram including all previously known planets and our {five} newly confirmed planets. The four candidates span a wide range of planetary densities. {TOI-5811\,B\,b has among the lowest density inferred of all known planets, but the grazing nature of the transit implies large uncertainties for its radius.} TOI-603\,b, TOI-2114\,b and TOI-5806\,b have densities compatible with those of Saturn, Jupiter and Neptune. {Finally,} TOI-4492\,b has among the largest bulk density in planets of similar mass ($\rho=6.37^{+0.49}_{-0.56}~$\gcm3), see also the middle panel in Fig.~\ref{fig:mass-radius}. As described in \cite{goswamy24} a possible explanation for such high-density gaseous giants is the presence of blended companions missed by the high-spatial resolution images and the RV analysis. In the case of TOI-4492, our high-spatial resolution images are able to discard companions down to 0.3\arcsec{} with contrasts of $\Delta m < 5$~mag in optical and near infrared bands. Additionally, the RVs do not show an evident slope, which in this case is compatible with zero ($-0.019\pm 0.055$~m/s/day) in a time span of 3.8 years of observations. For a companion to produce a significant increase in the planet radius (currently $1.051 \pm 0.055$~\Rjup) to move from the present density to a more typical density (below $3$~\gcm3), we would require a hidden stellar companion with a contrast magnitude smaller than $\Delta m < 0.5 $~mag (hence a similar brightness than the host star). Such companion should be located within $0.1\arcsec$ of the host star, as otherwise the high-spatial resolution images would have detected it. At the distance of TOI-4492 (142\,pc), 0.1\arcsec{} correspond to 14\,au. A similar-mass coplanar companion at such orbital distance would have produced a slope in the RVs of $+0.44$~m/s/day, which would have been detected by our RV dataset. Hence, we can safely discard the blended star scenario to explain the relatively anomalous high density of this planet. Instead, this high-density is compatible with being a consequence of the high mass of the planet ($5.92^{-0.67}_{-0.64}$~\Mjup). At such high masses, electron degeneracy dominates in hydrogen- and -helium-dominated bodies (e.g., \citealt{bashi17}) and consequently the radius slightly shrinks with increasing mass.

The case of TOI-603\,b is also remarkable, with properties placing this super-Neptune in the savanna, a scarcely populated region in the planet period-radius diagram defined by \cite{bourrier23} and likely sculpted by disk-driven migration processes. {The circular model is preferred over the eccentric model (with a 95\% upper limit on the eccentricity of $e_b<0.58$)}, in line with other known planets in this population, and in contrast with warm Jupiters, which have a broad range of eccentricities (see \citealt{correia20}). In the case of the hot-to-warm Jupiter in our sample, TOI-2114\,b, we do indeed find a moderately high eccentricity, of $0.472\pm0.026$. Such high eccentricities could be excited by either the presence of additional planets in the system gravitationally interacting, or by the remnant of a high-eccentricity migration process, proposed for the formation of hot Jupiters \citep[e.g., ][]{wu03}. Using Eq. (2) of \cite{adams03}, we find a typical tidal circularization time for this particular planet of 360 Myr (assuming $Q_P=10^6$). Our spectroscopic analysis can clearly discard such a young age for the system. Alternatively, there could be another source of eccentricity excitation in the system. One of the possible scenarios is an external planet. Our radial velocity dataset shows a hint of a long-term linear drift but still consistent with a zero slope ($-0.060^{+0.078}_{-0.077}$~m/s/day). Yet, the data lacks sufficient precision to establish firm conclusions on this scenario. Additional and higher precision RVs on this system are crucial to search for an hypothetical external planetary companion to account for the high eccentricity tentatively inferred in this work. Alternatively, star-planet interactions have also been proposed as a source for the excitation of eccentricities in hot-Jupiters (e.g., the elliptical instability proposed by \citealt{cebron13})

The transmission spectroscopy metric (TSM, \citealt{kempton18}) for each planet is ${\rm TSM}=91$ (TOI-603\,b), ${\rm TSM}=124$ (TOI-2114\,b), ${\rm TSM}=20$ (TOI-4492\,b), ${\rm TSM}=181$ (TOI-5806\,b), {and ${\rm TSM}=91$ (TOI-5811\,B\,b)}. TOI-603, TOI-2114\,b and TOI-5806\,b are excellent targets for atmospheric studies with the James Webb Space Telescope (JWST), with TSM values exceeding the typical 90 cutoff for high-quality atmospheric observations \citep{kempton18}. TOI-4492\,b on its side seems very compact, challenging any JWST attempt to detect its atmosphere. 

Besides, the confirmed planets orbit stars typically bright (T$_{\rm mag}=8-9$~mag) and two of them (TOI-2114 and TOI-5806) displaying broadened rotational profiles indicating moderately rapid rotation ($v\sin{i}$ being 10.7 and 19.1 km/s, respectively). With the planets being relatively large, this implies typical amplitudes for the Rossiter-McLaughlin effect at the level of 12-45 m/s assuming spin-orbit alignment. Consequently, these are ideal targets for obliquity measurements, key given the large eccentricity in the case of TOI-2114\,b and the presence of a close stellar companion in the case of TOI-5806\,b.

%________________________________________________________________
\section{Conclusions}
\label{sec:conclusions}

In this paper we present the first results of a follow-up program with the \cafe{} instrument on hot-Jupiter planet candidates from the TESS mission. These planets have been typically left behind in the effort to measure their masses and confirm their planetary nature, mainly due to their larger false positive rates (e.g., \citealt{santerne12}) but also due to the higher priority low-mass planets. Still, several works have focused on this type of planets \citep[e.g., ][]{rodriguez19,rodriguez21,rodriguez23,yee22,yee23,yee25} contributing to produce a growing catalog of hot- and warm- Jupiters critical to continue understanding the formation and atmospheric properties of these rare planetary configurations. We present five new planets in this regime (one of them is actually a super-Neptune in the savanna), one of them showing a dynamically interesting high eccentricity. Additionally, we reject the planetary candidate TOI-1837.01 as being a clear eclipsing binary. For TOI-1137.01, we present the only two possible scenarios matching the data, one including a planet orbiting the hidden companion star surrounding the RGB star and the other one including a brown dwarf. Finally, {TOI-5811 is a complex hierarchical triple system, A(ab)+B composed of three likely equal mass stars. Our data show that the transiting object is actually a planet around component B.} 
The results of this paper show the great capabilities of the \cafe{} instrument at Calar Alto observatory, demonstrating its ability to routinely study and confirm targets from the hot-Jupiter population.

% WARNING
%-------------------------------------------------------------------
% Please note that we have included the references to the file aa.dem in
% order to compile it, but we ask you to:
%
% - use BibTeX with the regular commands:
%   \bibliographystyle{aa} % style aa.bst
%   \bibliography{Yourfile} % your references Yourfile.bib
%
% - join the .bib files when you upload your source files
%-------------------------------------------------------------------

\bibliographystyle{aa} % style aa.bst
\bibliography{biblio2} % your references Yourfile.bib

\begin{appendix}
%\onecolumn
\nolinenumbers

\begin{acknowledgements}
% Referee
We thank the referee for their thorough and altruistic revision of this manuscript. 
% Facilities
We thank Calar Alto Observatory for allocation of director's discretionary time to this programme (DDT.26A.352).
This article is based on observations made with the MuSCAT2 instrument, developed by ABC, at Telescopio Carlos S\'anchez operated on the island of Tenerife by the IAC in the Spanish Observatorio del Teide.
 This research has made use of the Exoplanet Follow-up Observation Program (ExoFOP; DOI: 10.26134/ExoFOP5) website, which is operated by the California Institute of Technology, under contract with the National Aeronautics and Space Administration under the Exoplanet Exploration Program. Based on observations obtained at the Hale Telescope, Palomar Observatory, as part of a collaborative agreement between the Caltech Optical Observatories and the Jet Propulsion Laboratory operated by Caltech for NASA. Some of the data presented herein were obtained at Keck Observatory, which is a private 501(c)3 non-profit organization operated as a scientific partnership among the California Institute of Technology, the University of California, and the National Aeronautics and Space Administration. The Observatory was made possible by the generous financial support of the W. M. Keck Foundation. The authors wish to recognize and acknowledge the very significant cultural role and reverence that the summit of Maunakea has always had within the Native Hawaiian community. We are most fortunate to have the opportunity to conduct observations from this mountain. 
 This work makes use of observations from the LCOGT network. Part of the LCOGT telescope time was granted by NOIRLab through the Mid-Scale Innovations Program (MSIP). MSIP is funded by NSF.
 This research has made use of the Exoplanet Follow-up Observation Program (ExoFOP; DOI: 10.26134/ExoFOP5) website, which is operated by the California Institute of Technology, under contract with the National Aeronautics and Space Administration under the Exoplanet Exploration Program.
 Funding for the TESS mission is provided by NASA's Science Mission Directorate. KAC acknowledges support from the TESS mission via subaward s3449 from MIT and NASA grants 80NSSC24K1889 and 80NSSC26K0081.
% Personal acknowledgements
J.L.-B. is funded by the Spanish Ministry of Science and Universities (MICIU/AEI/10.13039/501100011033) and NextGenerationEU/PRTR grants PID2019-107061GB-C61, CNS2023-144309, and PID2023-150468NB-I00. B. M. acknowledges the funding by grant PID2021-127289-NB-I00 from MCIN/AEI/10.13039/501100011033/ and FEDER.
C.C. acknowledges financial support by the Consejo Superior de Investigaciones Cient\'ificas (CSIC) through the internal project 2023AT003 associated to the RYC2021-031640-I.
This work is partly supported by JSPS KAKENHI Grant Numbers
JP24H00017, JP24K17083, JP25K24620, JP26H01402, and JP26K00755 JSPS Bilateral Program Number JPJSBP120249910, and JSPS Grant-in-Aid for JSPS Fellows Grant Number JP24KJ0241.
F. M. acknowledges the financial support from the Agencia Estatal de Investigaci\'{o}n del Ministerio de Ciencia, Innovaci\'{o}n y Universidades (MCIU/AEI) through grant PID2023-152906NA-I00.
DRC acknowledges partial support from NASA Grant 18-2XRP18_2-0007.
A.C. acknowledges support from the National Aeronautics and Space Administration (80NSSC24K0496).
% Packages:
This work made use of \texttt{tpfplotter} by J. Lillo-Box (publicly available in \url{https://github.com/jlillo/tpfplotter}), which also made use of the python packages \texttt{astropy}, \texttt{lightkurve}, \texttt{matplotlib} and \texttt{numpy}.
This work made use of \texttt{TESS-cont} (\url{https://github.com/castro-gzlz/TESS-cont}), which also made use of \texttt{tpfplotter} and \texttt{TESS-PRF} \citep{bell22}.
This research made use of \texttt{Lightkurve}, a Python package for Kepler and TESS data analysis \citep{lightkurve}.
\end{acknowledgements}

% =======================================
\section{Ground-based photometry}
% =======================================
\label{app:GroundBasedPhotometry}
\FloatBarrier

We explored the ground-based seeing-limited photometric follow-up observations of our targets publicly available at ExoFOP. In Sect.~\ref{app:GBobservations} we describe these observations and in Sect.~\ref{app:GBresults} we present the results for per target.. 

\subsection{Observations}
\label{app:GBobservations}
% ииииииииииииииииии
\paragraph{MusCAT2 --} Data were collected for TOI-1137 (13-oct-2019), TOI-1837 (13-may-2020), and TOI-2114 (03-may-2025) with the multi-band imager MuSCAT2 \citep{narita19} mounted on the 1.5~m Telescopio Carlos S\'{a}nchez (TCS) at Teide Observatory, Spain. MuSCAT2 is equipped with four CCDs and can obtain simultaneous images in $g'$, $r'$, $i'$, and $z_s$ bands with negligible read-out time. Each CCD has $1024 \times 1024$ pixels with a field of view of $7.4 \times 7.4$~arcmin$^2$. The observations of the three targets were performed with the telescope defocused. For TOI-1137, the exposure times were set to 5~s for all bands; for TOI-1837, the exposure times were set to 6~s for $g'$, $i'$, and $z_s$, and 3~s for the $r'$ band; for TOI-2114, the exposure times were 11~s for $g'$ and $r'$, and 15~s for the $i'$ and $z_s$ bands. The raw data were reduced using the MuSCAT2 pipeline \citep{parviainen19}. The pipeline performs dark and flat-field calibrations, aperture photometry, and transit model fitting, including accounting for instrumental systematics.

% ииииииииииииииииии
\paragraph{LCO --} We observed TOI-1137 (16-June-2025, 18-Aug-2025), TOI-5806 (17-Oct-2022, 10-June-2023, 09-July-2024), TOI-1719 (12-Feb-2021), and TOI-5811Aab (10-June-2023, 11-Aug-2023, 06-Oct-2023) using the Las Cumbres Observatory Global Telescope \citep[LCOGT;][]{Brown:2013} 1.0\,m network nodes. We used the {\tt TESS Transit Finder}, which is a customized version of the {\tt Tapir} software package \citep{Jensen:2013}, to auto-schedule our transit observations. The 1.0\,m telescopes are equipped with a $4096\times4096$ SINISTRO camera having an image scale of $0\farcs389$ per pixel, resulting in a $26\arcmin\times26\arcmin$ field of view. The images were calibrated by the standard LCOGT {\tt BANZAI} pipeline \citep{McCully:2018}, and differential photometric data were extracted using {\tt AstroImageJ} \citep{Collins:2017}.

% ииииииииииииииииии
\paragraph{ULMT --}  We observed a full transit of TOI-1137.01 on 03-Oct-2024 in Sloan $i'$ band and on 08-Oct-2024 in Sloan $r'$ band from the 0.61\,m University of Louisville Manner Telescope (ULMT). ULMT is located at the Steward observatory near Tucson, AZ. The $9576\times6388$ ASI 6200 camera has an image scale of $0\farcs165$ per pixel, resulting in a $26\arcmin\times17\arcmin$ field of view. The images were calibrated and photometric data were extracted using {\tt AstroImageJ}.

% ииииииииииииииииии
\paragraph{KeplerCam --} We observed TOI-2114.01 on 18-may-2021, 18-June-2021, and 07-April-2023 in Sloan $i'$ from KeplerCam, which is installed on the 1.2\,m telescope at the Fred Lawrence Whipple Observatory. The $4096\times4096$ Fairchild CCD 486 detector has an image scale of $0\farcs672$ per $2\times2$ binned pixel, resulting in a $23\farcm1\times23\farcm1$ field of view. The images were calibrated and photometric data were extracted with {\tt AstroImageJ}.

% ииииииииииииииииии
\paragraph{GMU --} We observed TOI-2114.01 in R-band on 17-May-2021 from the George Mason University 0.8\,m telescope near Fairfax, VA. The telescope is equipped with a $4096\times4096$ SBIG-16803 camera having an image scale of $0\farcs35$ per pixel, resulting in a $23\arcmin\times23\arcmin$ field of view. The images were calibrated and photometric data were extracted using {\tt AstroImageJ}.

% ииииииииииииииииии
\paragraph{Novosibirsk --} We observed TOI-4492 on 04-Sep-2025 in Sloan $g'$ and Sloan $i'$ from the Lyceum 130 Observatory 0.28\,m telescope near Novosibirsk, Russia. The telescope is equipped with a $4176\times6248$ pixels ZWO ASI2600MM camera having an image scale of $0.28~\arcsec$/pixel, resulting in a $19\arcmin\times29\arcmin$ field of view. The images were calibrated and photometric data were extracted using {\tt AstroImageJ}.

% ииииииииииииииииии
\paragraph{OAUV --} We observed TOI-4492.01 on 07-Apr-2022 from the Observatori Astron\`omic de la Universitat de Val\`encia (OAUV, Val\`encia, Spain). The 0.14\,m telescope is equipped with a FLI ML16200 detector that has a image scale of $3.1$~$\arcsec/$pixel, resulting in a $233\arcmin\times186\arcmin$ field of view. The images were calibrated and photometric data were extracted using {\tt AstroImageJ}.

% ииииииииииииииииии
\paragraph{SUTO --} 
We observed TOI-4492.01 in R band on 17_july-2025 from the Silesian University of Technology (SUTO) 0.3\,m telescope, which is located at near Otivar, Spain. The telescope is equipped with a $4656\times3520$ pixel ASI ZWO 1600MM detector that has an image scale of 0$\farcs$685 pixel$^{-1}$, resulting in a $53\arcmin\times42\arcmin$ field of view. The differential photometric data were extracted using {\tt AstroImageJ}.

% ииииииииииииииииии
\paragraph{Acton Sky Portal --} 
We observed TOI-1137.01 in Sloan $r'$ on 06-Oct-2019 from the Acton Sky Portal private observatory in Acton, MA, USA. The 0.36\,m telescope is equipped with an SBIG-ST8XME camera having an image scale of $0.69$\arcsec/pixel, resulting in a $17\arcmin{}\times11.5\arcmin$ field of view. The image data were calibrated and photometric data were extracted using {\tt AstroImageJ}.

\subsection{Results per target}
\label{app:GBresults}

%----------------------------------------
\subsubsection{TOI-1137}
%----------------------------------------
\label{app:TOI-1137}

For the transit observation of TOI-1137.01, a reanalysis of the data (compared to the results available at ExoFOP) using the most updated ephemeris predicts that an ingress should have been detected within the observed window of the MusCAT2 observations. However, no significant signal is found. Although no transit is detected on the target, the data rules out all nearby stars detected in \gaia{} DR3 as potential eclipsing binaries (EBs) within the TESS pixel area. The additional observations from LCO and ULMT are consistent with a $\sim$2 ppt event in the z-band (two transits from LCO) and in i$_p$-band (two transits from ULMT). One additional B-band observation from LCO is consistent with no detection after airmass detrending. 

%----------------------------------------
\subsubsection{TOI-1837}
%----------------------------------------
\label{app:TOI-1837}
For the transit observation of TOI-1837.01, no significant transit detection was achieved on the MuSCAT2 observations, showing a photometric scatter comparable to the expected transit depth. Several neighboring stars can be ruled out as EBs during the observed window. However, these observations cannot resolve the close stellar companion reported in the high-spatial resolution images and are thus not useful to unveil the source of the eclipses. 

%----------------------------------------
\subsubsection{TOI-2114}
%----------------------------------------
\label{app:TOI-2114}
For TOI-2114.01, MuSCAT2 detected a transit event on the target star. The transit depth is consistent across all bands within uncertainties, although the $z_s$ band appears slightly shallower than the others. This could be attributed to differences in the treatment of limb darkening (as the transit fit was performed without imposing priors on limb darkening coefficients) or imperfect modeling of systematic effects present in the time series. We thus conclude that the transit is achromatic in the optical bands, reinforcing the planetary origin of the transiting object. Additional observations from KeplerCam and GMU show partial transits on target also compatible with the TESS transit depth. 

%----------------------------------------
\subsubsection{TOI-4492}
%----------------------------------------
\label{app:TOI-4492}

The Novosibirsk observations detected a transit in both Sloan $g'$ and Sloan $i'$ bands with depth consistent with the TESS data, thus confirming no strong transit depth chromaticity. SUTO also detects a full transit in R band with depth consistent with TESS, after detrending a mid-transit baseline shift caused by a telescope meridian flip. The OAUV observations only covered a short in-transit segment and egress, but with airmass detrending, the data show an egress with depth consistent with TESS.

%----------------------------------------
\subsubsection{TOI-5806}
%----------------------------------------
\label{app:TOI-5806}
There are three LCO observations in zs band from 2021, 2023, and 2024. The 2023 and 2024 observations have good enough focus and seeing to separate the 2.1\arcsec{} neighbor and demonstrate that the event is indeed on target.

%----------------------------------------
\subsubsection{TOI-5811}
%----------------------------------------
\label{app:TOI-5811}
The LCO observation of TOI-5811A(ab) obtained on 10-jul-2023 caught about 90\% of the transit window, according to the QLP ephemeris from sectors 55 and 82. On those observations, we find a $\sim$2~ppt event on component {B} of this triple system  (TIC 331484427, $\Delta T = 0.561$~mag, 7.46\arcsec{} NW). The light curve from the main target (composed of stars {Aa and Ab}) show no clear events on this time window. However, due to the lack of pre-ingress baseline, these results are only tentative and non-conclusive, although they seem to point to star {B} as the origin of the transits. {This hypothesis is actually confirmed by our CARMENES and TRES data (See Sect.~\ref{sec:analysis})}.

% ==================================================================
\section{Analysis of the nature of the TOI-1137 transiting signal}
% ==================================================================
\label{app:TOI-1137}
\FloatBarrier

In Sect.~\ref{sec:TOI-1137}, we argued that the short orbital period reported for TOI-1137.01 reject the hypothesis of this object transiting the TOI-1137 (component A) given its evolved stage (with the radius of the star being larger than the orbital semi-major axis). Here, we report on the additional knowledge in hand to discuss about the two possible scenarios for C: 1) a low-mass or brown dwarf transiting star B, or 2) a planet transiting star B.  

Our \cafe{} spectra (clearly dominated by light from the giant star) display a RV scatter at the 25 m/s level and no significant correlation between the RV and the FWHM over the two year time span of the observations. Therefore, this suggests that star B is either very faint, or it is bright but has very broad lines due to rapid rotation ($v\sin{i}>30$km/s), either of which would prevent star B from affecting the RV measurements.

The TESS light curve, however, shows a transit depth of $\sim 1050$~ppm. The maximum contrast that an EB could have to produce such a transit depth ($\delta$) is $\Delta m^{\rm max} = -2.5\log_{10}{\delta}=7.5$~mag. Hence, under the assumption that both the evolved star and the companion have the same age (i.e., they are bound and coetaneous), the spectral type of the companion must be M1~V or earlier. As such, given the high-spatial resolution imaging sensitivity limits, it must be separated from the evolved star by less than $0.5\arcsec$ (corresponding to $\sim$200 au at the distance of TOI-1137 obtained from \gaia{} DR3). 
Interestingly, the undiluted depth of the transit, given by $\delta^{\rm true}=\delta^{\rm obs}\times (1+10^{-\Delta m/2.5})$, and implying a corresponding radius of the transiting object given by $R_p=R_{\star}\sqrt{\delta^{\rm true}}$, would only stay in the planetary/sub-stellar domain ($R_p\lessapprox 2.5$~\Rjup) if the companion star has $\Delta m < 5$ mag in the TESS band. This corresponds to a spectral type earlier than K3. 
In this scenario, the flux ratio between B and A would be of the order of $10^{-2}$ or higher. At the lower limit, B would hardly be visible in the \cafe{} spectrum. Indeed, the standard deviation of the residual CCF after the Gaussian modeling in our \cafe{} data represents 0.9\% of the contrast of the CCF. This implies that we would detect a companion with a contrast depth of 2.7\%. By approximating this to the flux ratio between components A and B, $F_B/F_A<0.027$, this corresponds to a K0V star, marking this as the earlier type for a potential bound companion to be missed in the \cafe{} CCF. 

Hence, in the planet hypothesis, the companion star must be bound to the evolved star and be in the K0-K3 dwarf regime. Such a star would be 4.1-5.0 mag fainter than component A in the V band. According to the sensitivity limit of the 'Alopeke high-spatial resolution image in the 562\,nm band, this implies that the companion should be closer than $0.1\arcsec$ to remain undetectable. This corresponds to $<$41~au.%, approximately between the orbits of Neptune and the aphelion of Pluto in our Solar System. 

\begin{figure}
	\centering
	\includegraphics[width=0.45\textwidth]{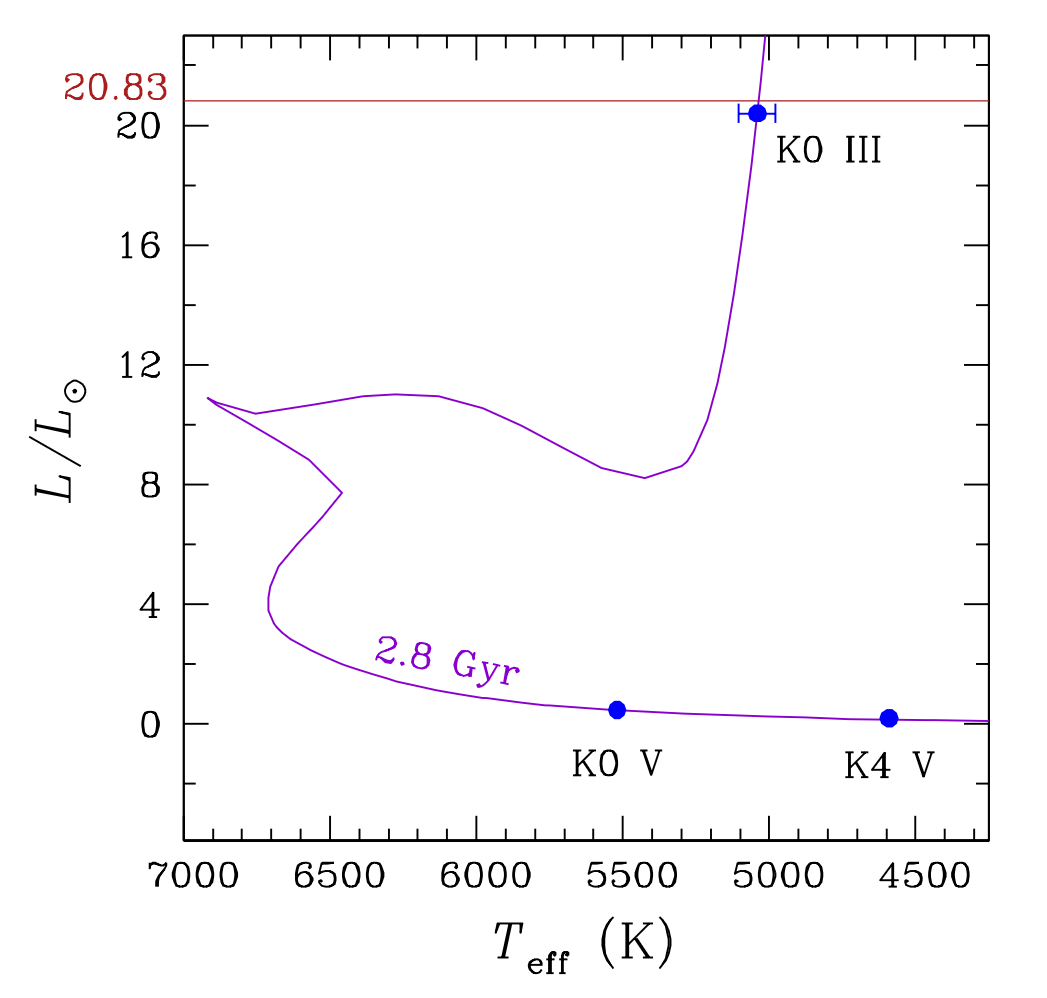}
	\caption{HR diagram for the binary system composed of a primary K0~III and two possibilities for the secondary, namely, K0 V or K4 V. The position of the giant and the two potential secondaries explored for the K dwarfs have been set so that $L$(K0 III)+$L$(K dwarf)=20.83 $L_\odot$. PARSEC 2.1 isochrones are used (blue dots.}
	\label{fig:TOI1137_HR}
\end{figure}

We cross-checked the viability and robustness of this scenario by proving that both stars could be coetaneous. The binary would be composed of a primary K0 III star and two extreme possibilities for the secondary, namely K0 V or K4 V. Figure~\ref{fig:TOI1137_HR} shows two HR diagrams for this configuration, which is compatible with the fact that the transiting object is a planet. The luminosities of the K dwarfs considered -individually- as potential companions are in the range 0.42 - 0.19 L$_{\odot}$ , respectively, according to the Schmidt-K\"aler compilation of standard data for stars \citep[][eds]{LB82}, therefore the bulk of the luminosity received from the object, 20.83 L$_{\odot}$, (brown horizontal line in the luminosity-temperature diagram), corresponds to the giant star. For this exercise, the luminosity of the K0 III has been set to 20.4 L$_{\odot}$, so the total luminosity $L$(K0 III)+$L$(K dwarf) matches the observed value. A temperature T = 5041 K for the giant has been used, and temperatures of 5520 and 4590 K have been assigned to the K0 V and K4 V, respectively \citep[][eds]{LB82}. The pairs K0 III + K dwarf -blue dots in both diagrams- are located on the same isochrone (2.8 Gyr). PARSEC 2.1 tracks and isochrones for metallicity [M/H]=$-0.35$ \citep{bressan12} have been used for this part of the work. The parameters of the giant, according to its position on the HR diagrams are $\log g\!=\!3.04$, 1.38 M$_{\odot}$ and 5.89 $R_{\odot}$. The red dot in the gravity - temperature diagram corresponds to an independent determination of this parameter through the analysis of the spectral energy distribution, adding confidence to the parameters derived in this exercise.

\begin{figure}[H]
	\centering
	\includegraphics[width=0.49\textwidth]{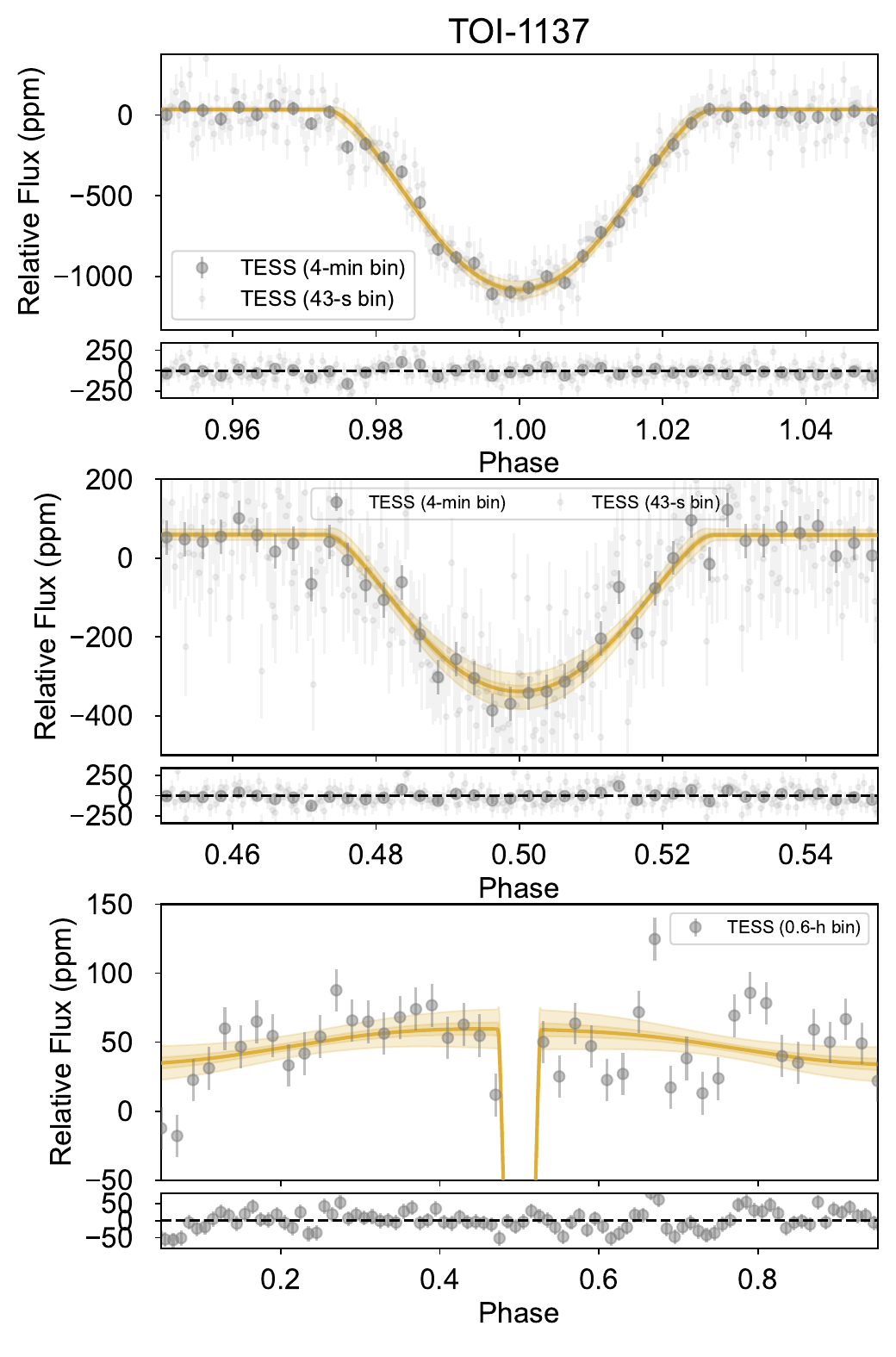}
	\caption{TESS light curve modeling for TOI-1137 assuming the eclipsing binary scenario. In all panels we show the the data with gray symbols (tone as indicated in the legend of each panel) and the model in gold color with the 68.7\% confidence interval as a shaded region of the same color. The residuals to this median model are also shown in the small panel below. \textbf{Top panel:} Zoom-in to the primary eclipse orbital phases. \textbf{Mid panel:} Zoom-in to the secondary eclipse orbital phases. \textbf{Bottom panel:} Zoom-in to the phase curve amplitude.}
	\label{fig:TOI1137_EB}
\end{figure}

The alternative to this planetary scenario is a blended EB sufficiently faint to escape the CCF detection ($\Delta m > 4$~mag, corresponding to a star later than K0~V, as explained above), but sufficiently bright to be able to mimic the observed transit depth (earlier than M1~V, as explained above). We model the TESS light curve under this scenario, assuming the eclipses are produced by another stellar or sub-stellar object (implying that the night and day temperatures are equal and correspond to the effective temperature of the object producing the eclipses). In this case, the host star (star B) is assumed to be a K0~V for simplicity\footnote{As demonstrated above, if star B is later than a K3~V, then the radius of the transiting object is still compatible with a planetary origin, although such radius limit only imposes a limit on the mass of around $< 200$~\Mjup{} (0.2\Msun) from \cite{baraffe03}.}. Even in this case, the modeling (see Fig.~\ref{fig:TOI1137_EB}) favors an scenario where the eclipsing object has a mass of $20.0^{+10.0}_{-7.2}$~\Mjup{} based on the detection of low-amplitude ellipsoidal variations (with observed non-dilution corrected amplitude of $A_{\rm ellip}=35^{+16}_{-19}$~ppm). We also infer an inflated radius of $2.35^{+0.26}_{-0.20}$~\Rjup, likely due to the proximity to both its parent host and the nearby RGB star (component A). The inferred temperature of the companion is in the 3000~K regime, hence potentially being compatible with an L-type brown dwarf. However, the detection of the ellipsoidal variations is not yet significant and we can thus not confirm the brown dwarf nature of the transiting object. Additional observations are required to confirm this signal. 

Consequently, with the data in hand we cannot determine whether the transiting object is a planet or a gravitationally bound EB. Our data could still be compatible with a planet around a K0-K3 star closer than $0.1\arcsec$. If the companion is brighter than $\Delta m< 4.1$ mag (with spectral types earlier than K0~V), then the only possibility for the planet scenario is that the star rotates so fast that its CCF is so broadened that it is not detectable in the CCF. Alternatively, if the companion is later than K3, then the undiluted radius of the transiting object would be in the stellar domain. 

% =======================================
\section{Additional figures}
% =======================================
\FloatBarrier
%_____________________________________________________________
%                        Figures
%-------------------------------------------------------------

\begin{figure}[h]
	\includegraphics[width=0.25\textwidth]{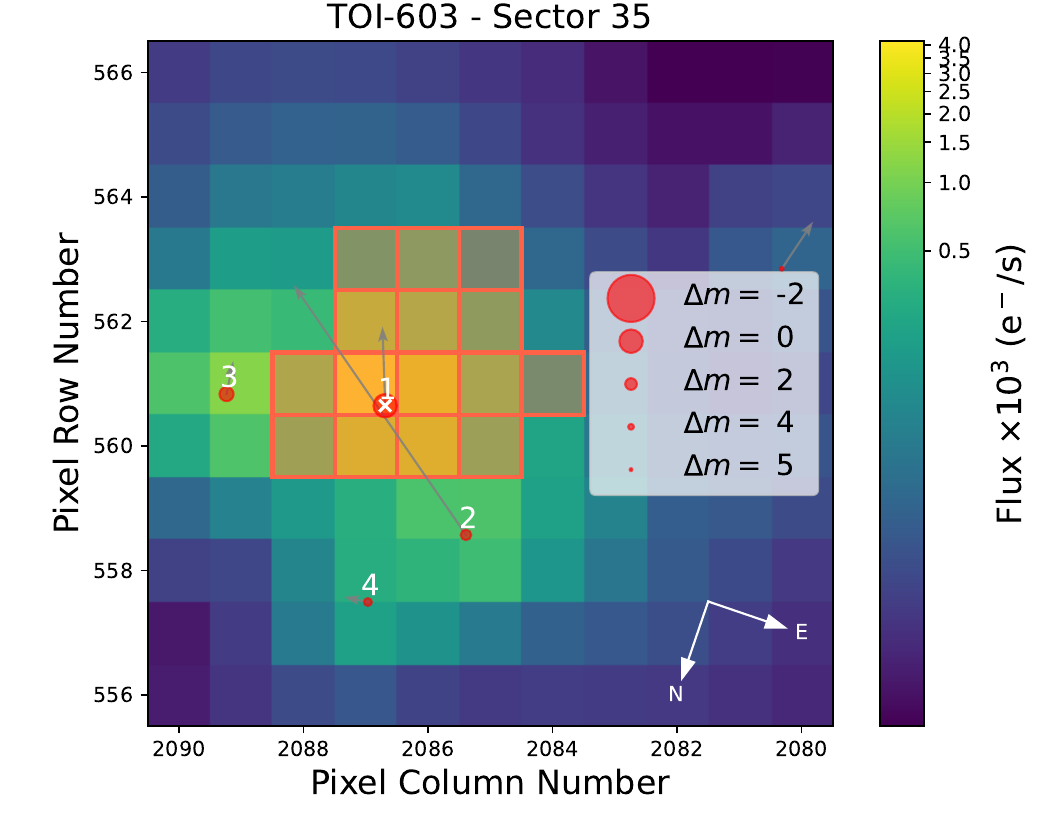}\includegraphics[width=0.25\textwidth]{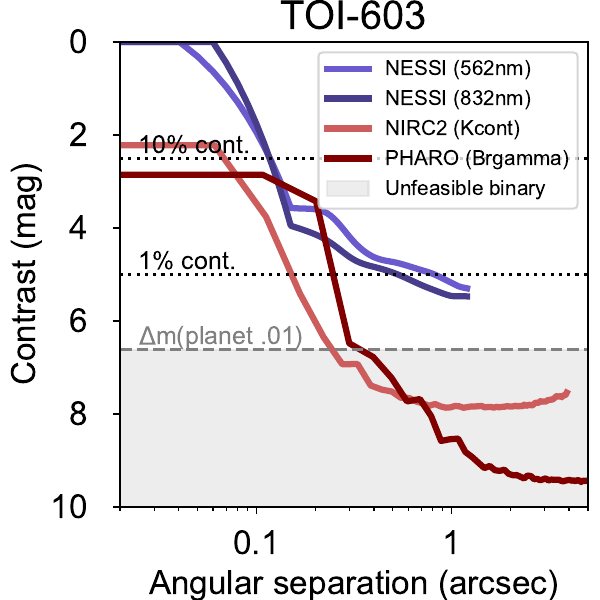}\vspace{0.5cm}
	
	\includegraphics[width=0.25\textwidth]{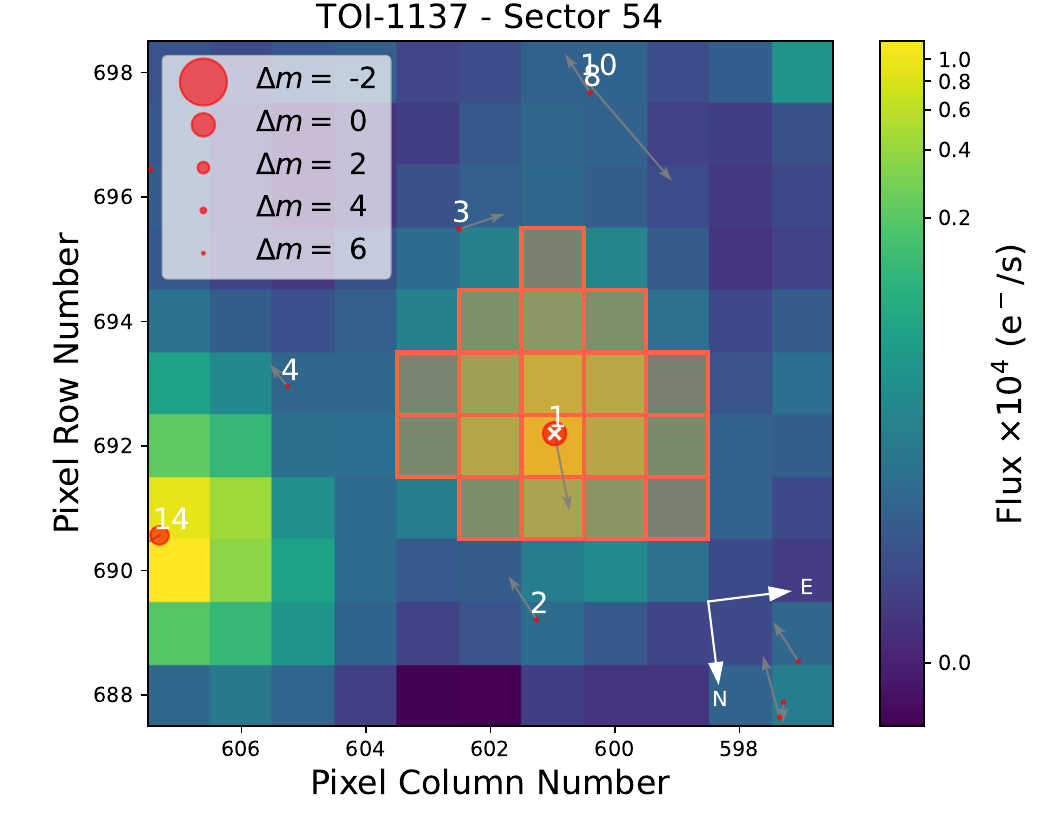}\includegraphics[width=0.25\textwidth]{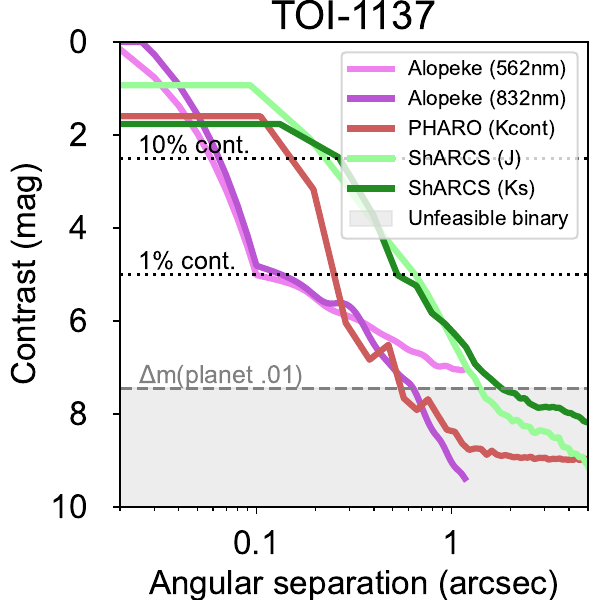}

   \caption{For each target (one per row), we present in the left panels the Target Pixel Files (TPF) plots using the \texttt{tpfplotter} algorithm \citep{aller20} and studied in Sect.~\ref{sec:TESS}, and the sensitivity curves from the high-spatial resolution images presented in Sec.~\ref{sec:HRimaging}.}
   \label{fig:tpfplots}
\end{figure}

\begin{figure}[h]

	\includegraphics[width=0.25\textwidth]{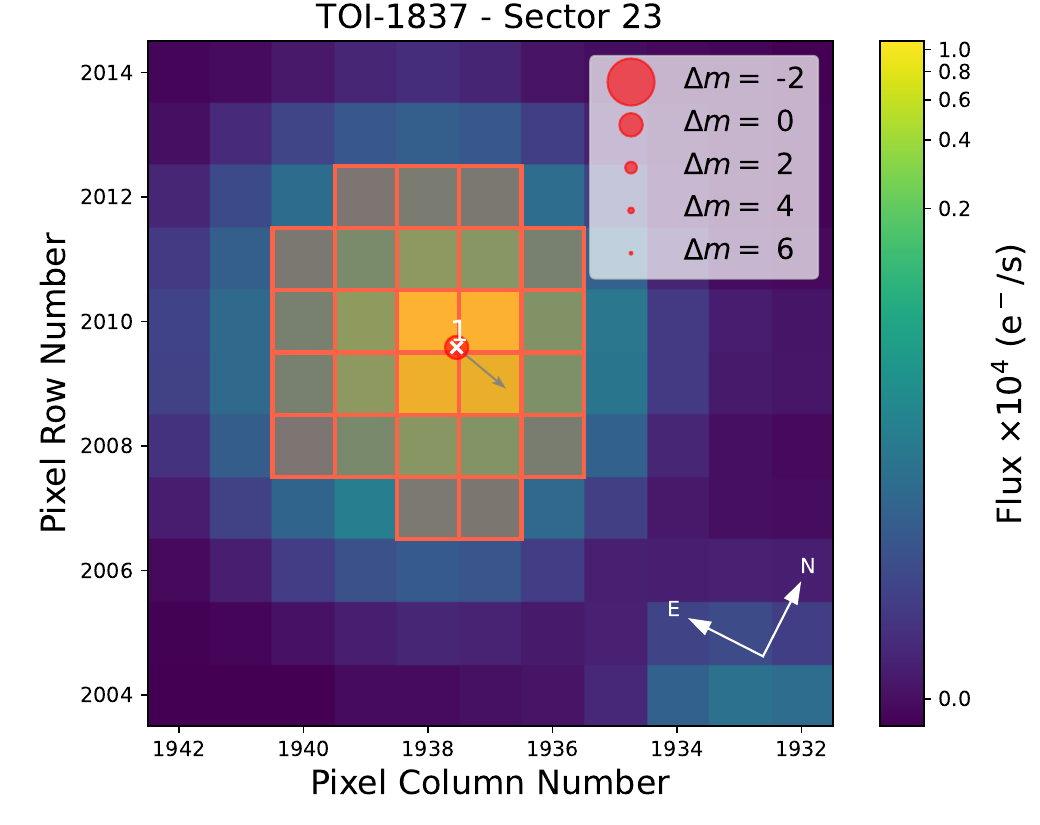}\includegraphics[width=0.25\textwidth]{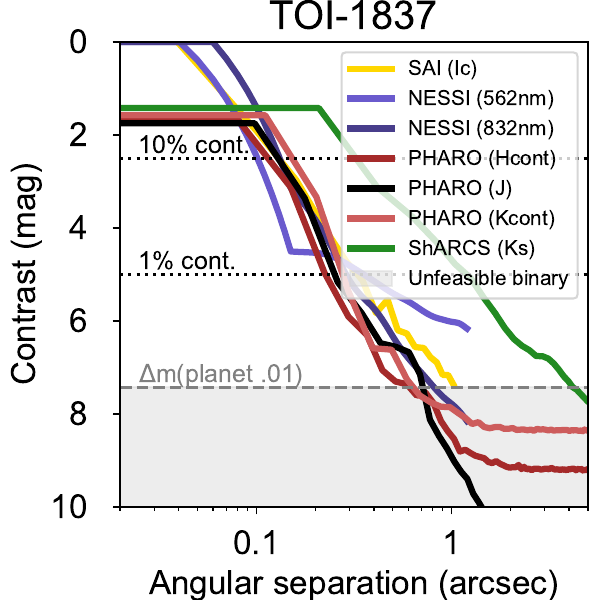}\vspace{0.5cm}
	
	\includegraphics[width=0.25\textwidth]{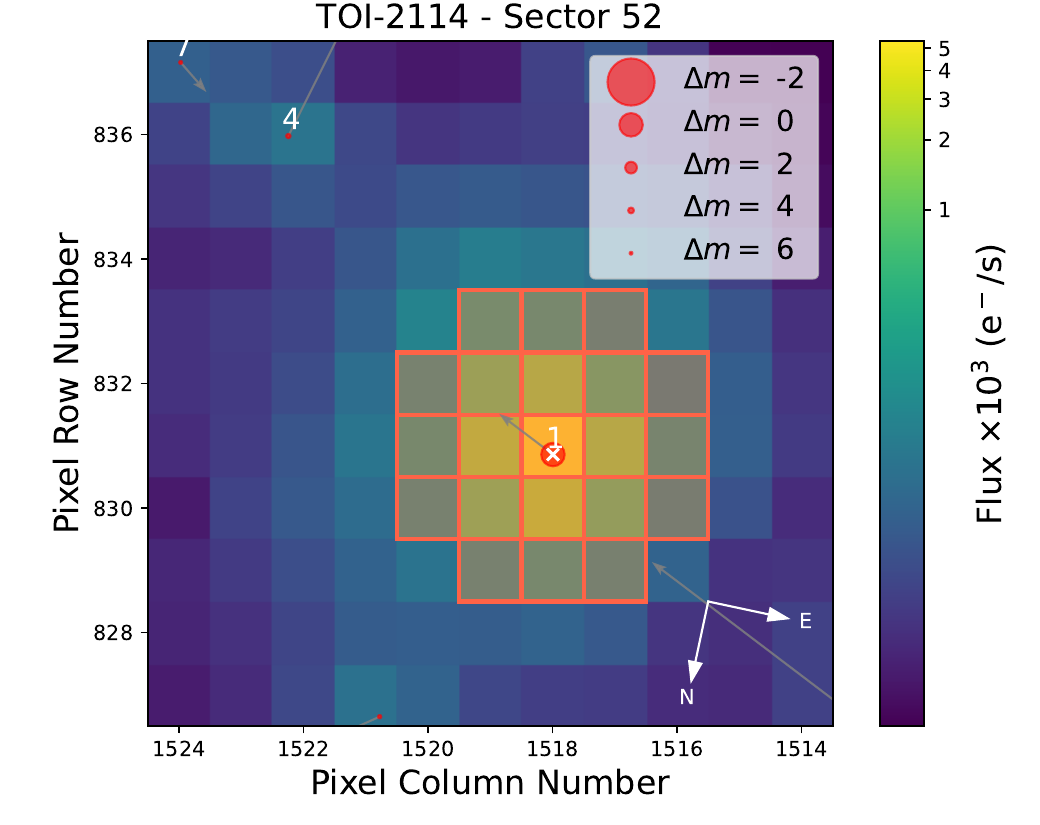}\includegraphics[width=0.25\textwidth]{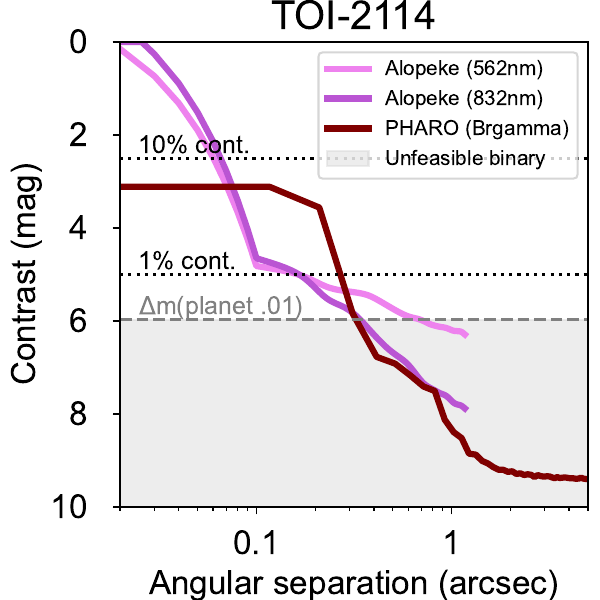}\vspace{0.5cm}
	
	\includegraphics[width=0.25\textwidth]{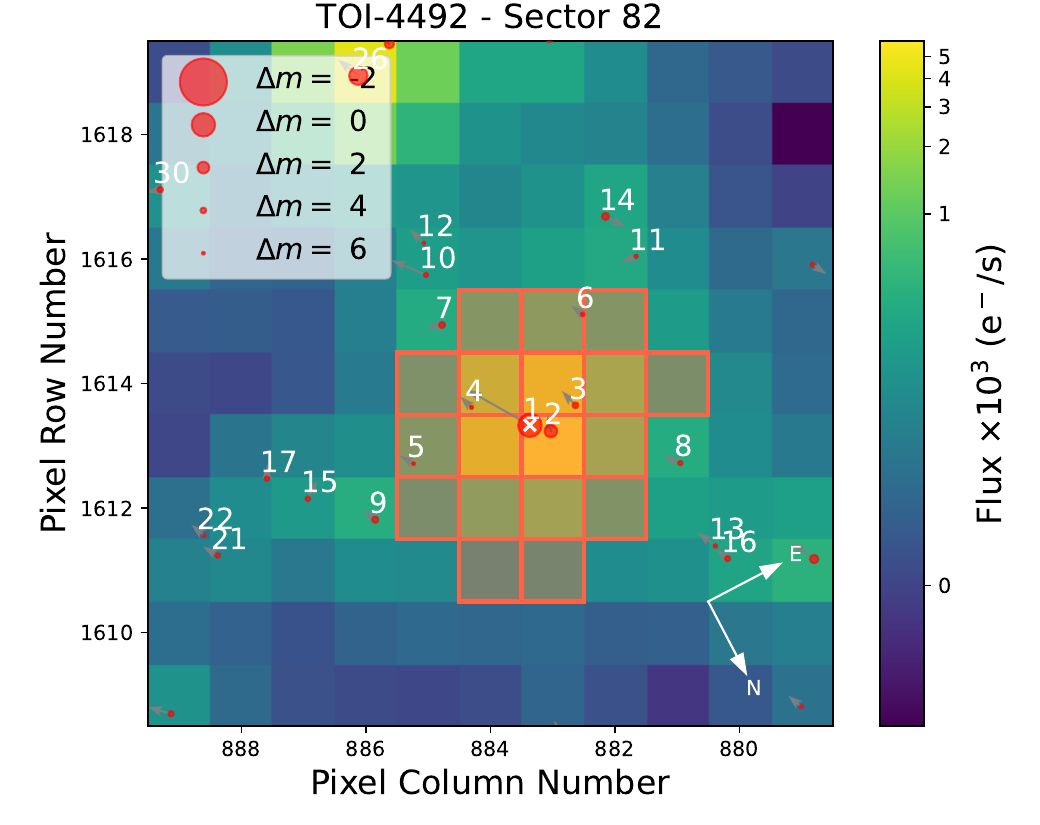}\includegraphics[width=0.25\textwidth]{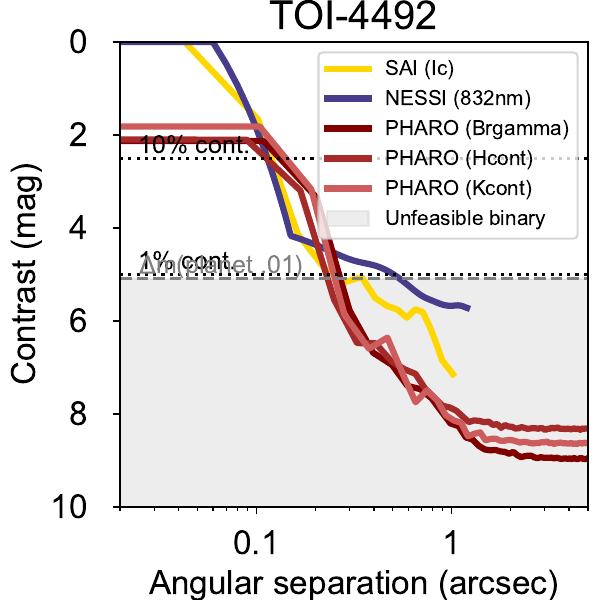}\vspace{0.5cm}
	
	\includegraphics[width=0.25\textwidth]{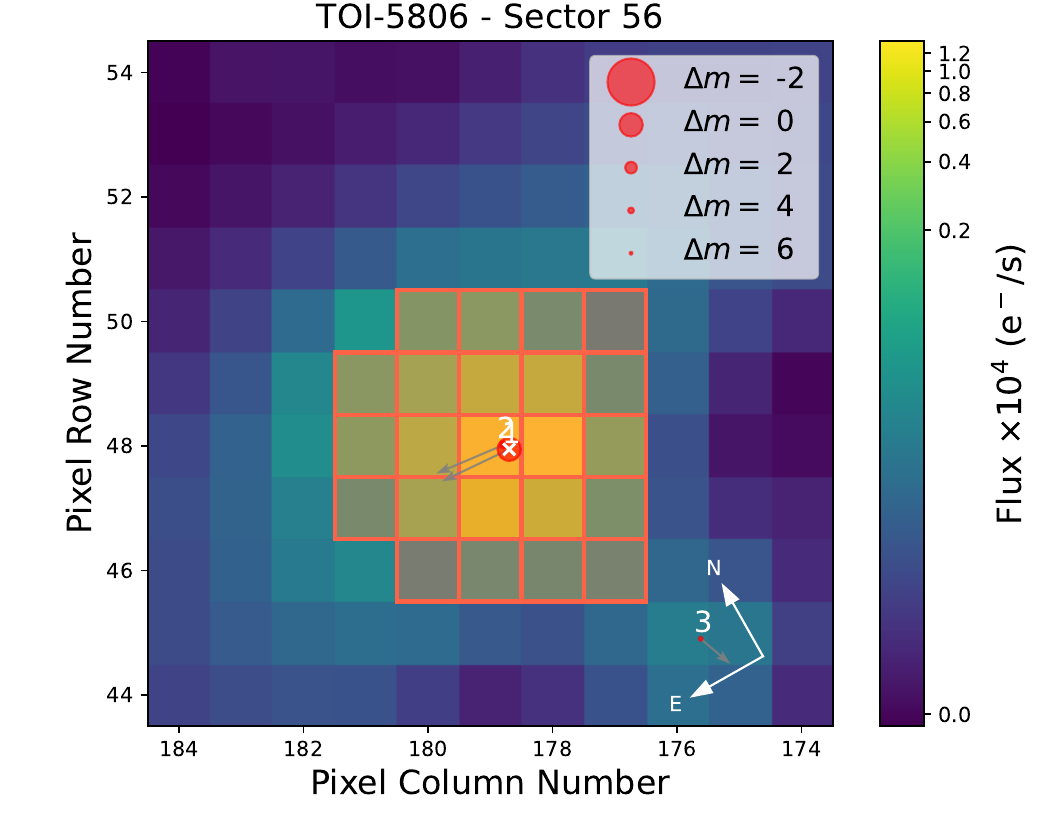}\includegraphics[width=0.25\textwidth]{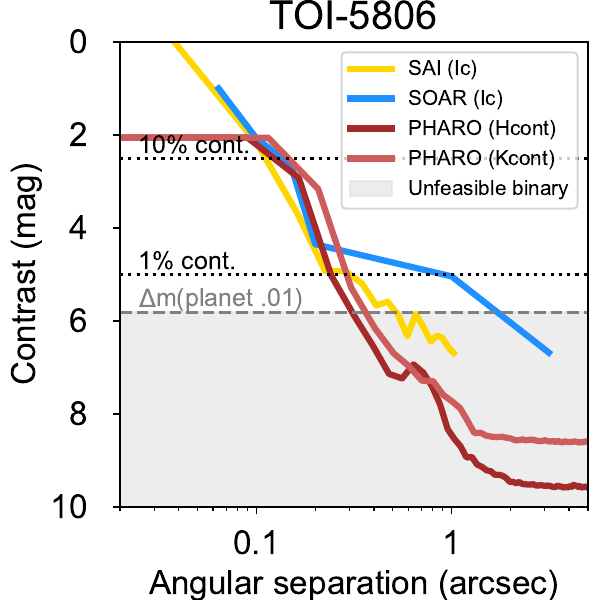}

   \caption{Continuation of Fig.~\ref{fig:tpfplots}.}
   \label{fig:tpfplots2}
\end{figure}

\begin{figure}
\centering
	\includegraphics[width=0.45\textwidth]{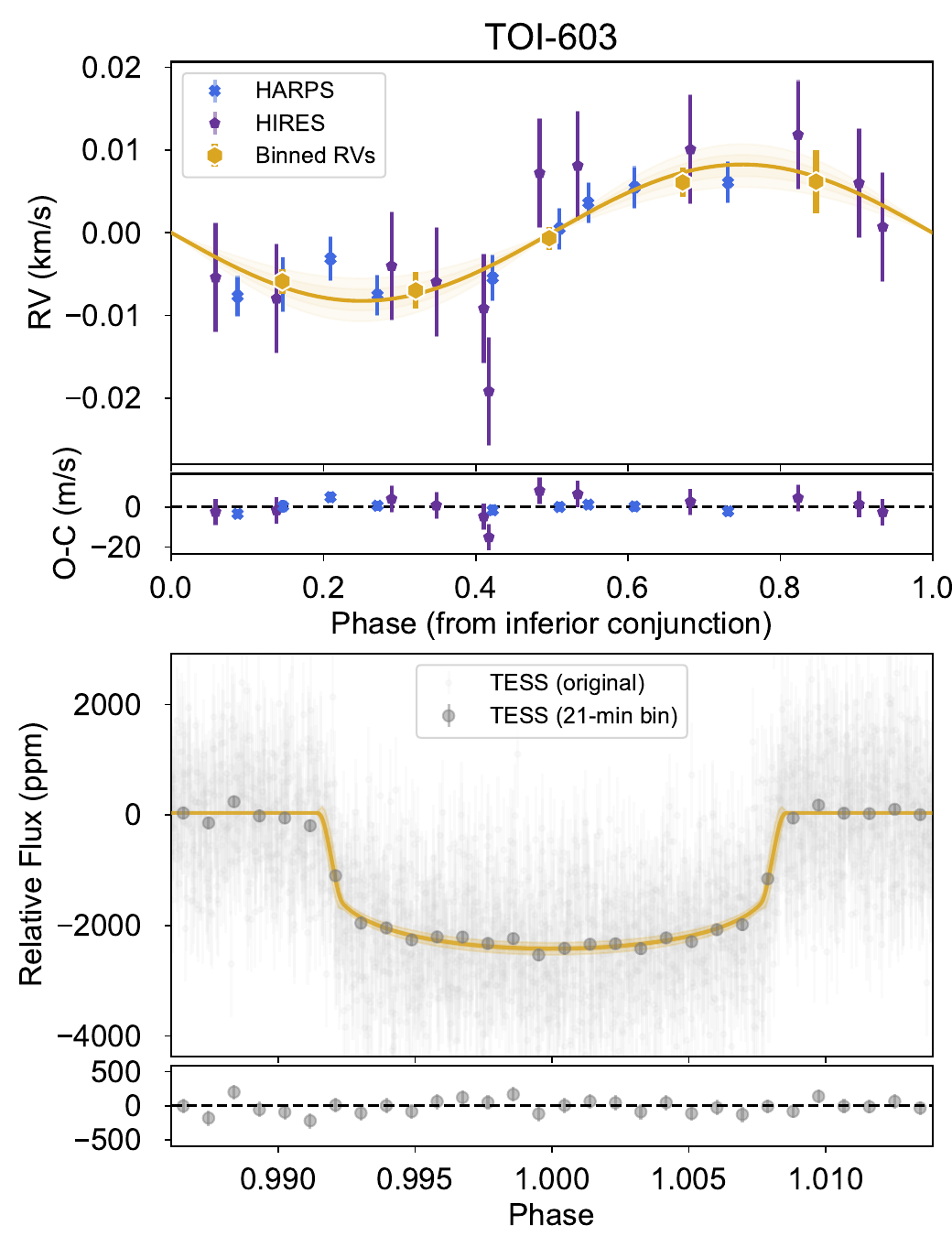}
	\caption{Results for TOI-603\,b from the joint radial velocity and light curve analysis. Panel descriptions are the same as Fig.~\ref{fig:result-TOI4492}.}
	\label{fig:result-TOI603}
\end{figure}	

\begin{figure}
	\centering
	\includegraphics[width=0.45\textwidth]{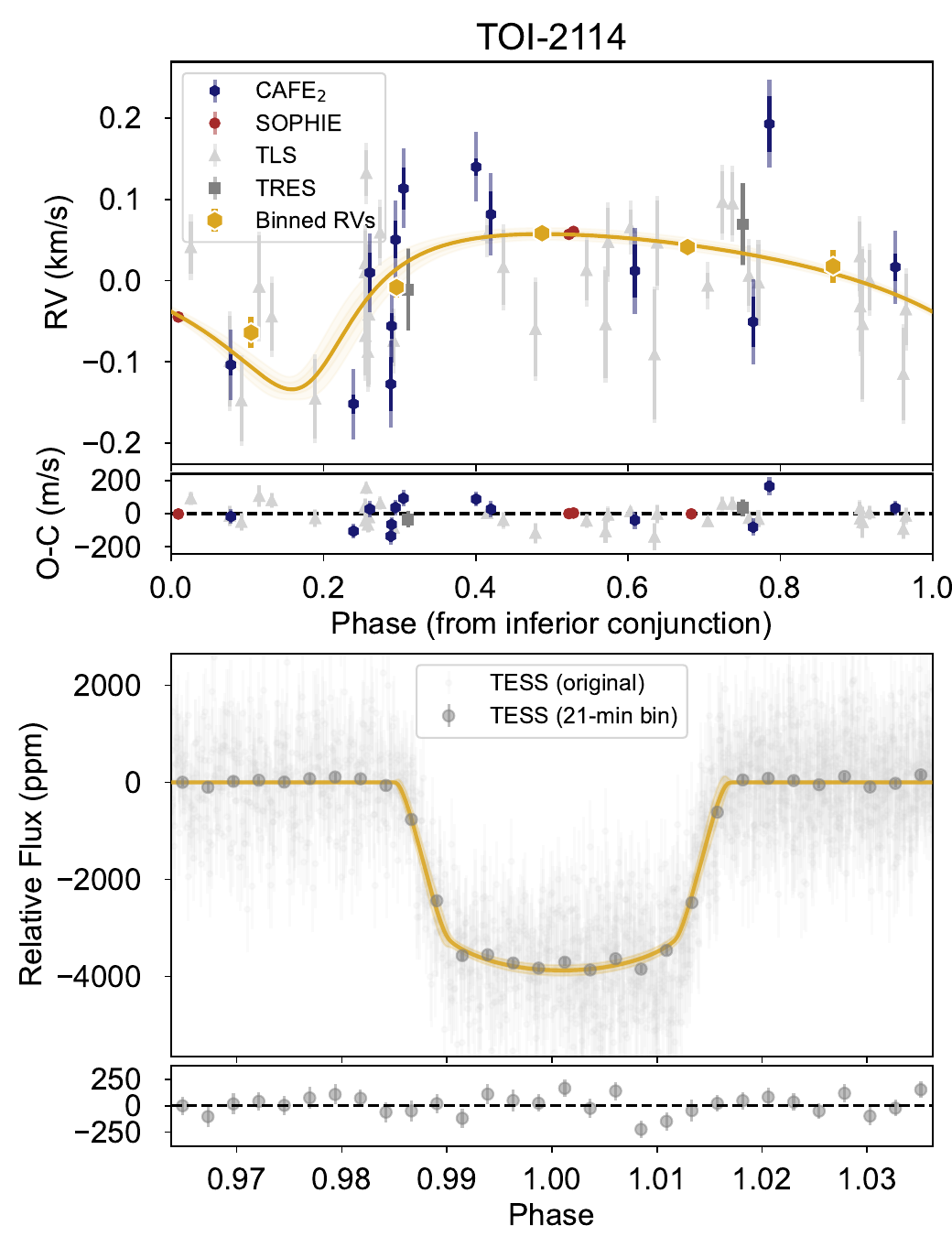}
	\caption{Results for TOI-2114\,b from the joint radial velocity and light curve analysis. Panel descriptions are the same as Fig.~\ref{fig:result-TOI4492}.}
	\label{fig:result-TOI2114}
\end{figure}    

\begin{figure}
	\centering
	\includegraphics[width=0.45\textwidth]{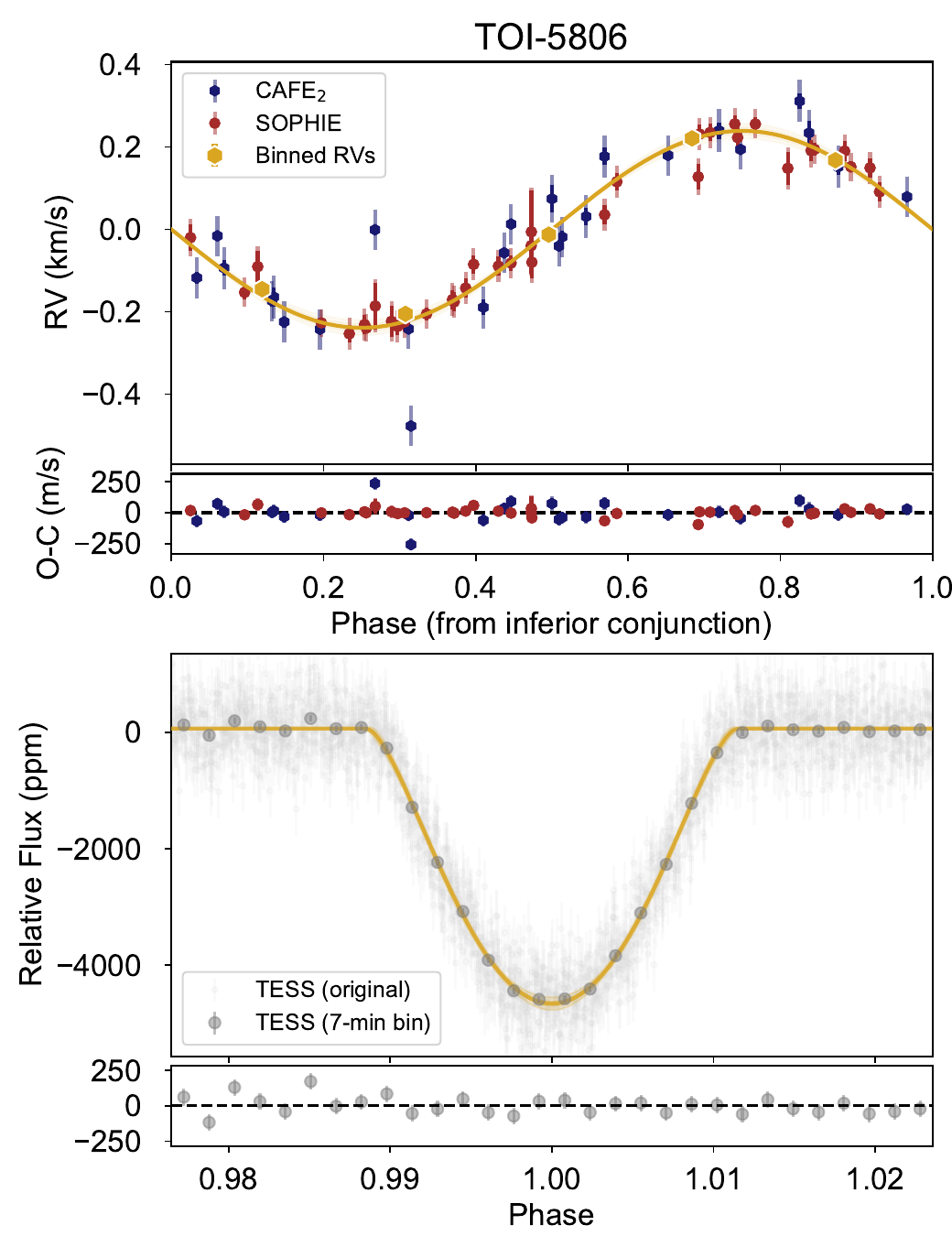}
	\caption{Results for TOI-5806\,b from the joint radial velocity and light curve analysis. Panel descriptions are the same as Fig.~\ref{fig:result-TOI4492}.}
	\label{fig:result-TO5806}
\end{figure}

\begin{figure}
	\centering
	\includegraphics[width=0.45\textwidth]{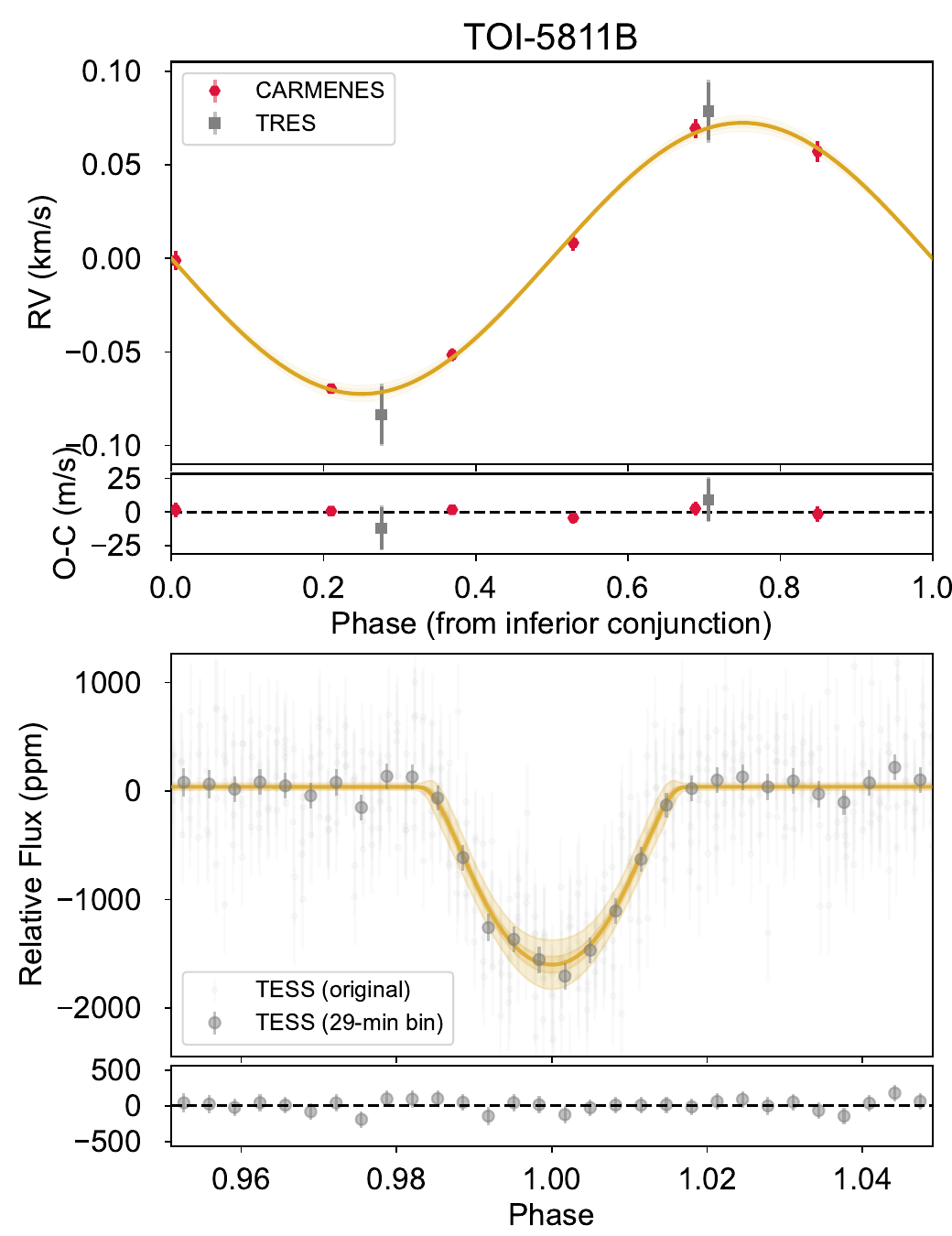}
	\caption{{Results for TOI-5811\,B\,b from the joint radial velocity and light curve analysis. Panel descriptions are the same as Fig.~\ref{fig:result-TOI4492}.}}
	\label{fig:result-TO5811}
\end{figure}

\FloatBarrier

% =======================================
\section{Additional tables}
% =======================================

%_____________________________________________________________
%                        TABLES
%-------------------------------------------------------------

%_____________________________________________________________
%
\begin{table}[h]
\setlength{\extrarowheight}{3pt}
\small
\caption{TESS and radial velocity observations used in this work for each of the targets.}             % title of Table
\label{tab:target_obs}      % is used to refer this table in the text
\begin{tabular}{r l c c c l }        % centered columns (4 columns)
\toprule
TOI  & Sectors & N$^{RV}_{\rm CAFE}$ & t$_{\rm span}$ & $\sigma_{\rm RV, CAFE}$ & N$^{RV}_{\rm others}$   \\
     &         &                     & (days)         & (m/s) &   \\    % table heading
\midrule
  603   & [8, 35, 45,         & 15  &  86  & 16.5  & 4 (TRES)   \\
           & 46, 61, 72]         &     &      &       &            \\
  1137  & [14, 54]                       & 9   &  282  & 6.5  & 2 (TRES)  \\
%  1719  & [20, 21, 47, 74]               & 49  &  472  & 6.4  & 2 (TRES)   \\
  1837  & [23,77]                        & 12  &  8  & 60     &   \\
  2114  & [25, 26, 52,           & 13  &  360  & 25   & 2 (TRES), \\
        &  53, 79]                              &     &       &      & 4 (SOPHIE), \\
        &                                &     &       &      & 35 (TLS)  \\
  4492  & [14, 15, 41,           & 31  &  365  & 9.2  & 2 (TRES), \\
          &  55, 82]                                &     &       &      &  6 (NEID)  \\
  5806  & [55\tablefootmark{a}, 56, 82]  & 25  &  359  & 13.4 &   36 (SOPHIE)\\
  5811\,A  & [55\tablefootmark{a}, 82]      & 29  &  366  & 9.2  & 2 (TRES)  \\
  5811\,B  & [55\tablefootmark{a}, 82\tablefootmark{a}]      & 0  &  -  & -  & 2 (TRES)  \\
          &     &     &       &      &  6 (CARMENES)  \\

\bottomrule
\end{tabular}
\tablefoot{
\tablefoottext{a}{This sector is only available in the QLP pipeline and is thus not used in this study.}
}
\end{table}
%_____________________________________________________________

%
\begin{table*}[h]
\caption{High-spatial resolution imaging results. In the cases where a close source is detected, we provide the projected angular separation ($\rho$), the contrast magnitude ($\Delta m$) in the corresponding band, and the position angle (PA). See Sect.~\ref{sec:HRimaging} for additional details. }             % title of Table
 \setlength{\extrarowheight}{3pt}
 \small
\label{tab:HRimaging}      % is used to refer this table in the text
\begin{tabular}{l l l l l l l l l}        % centered columns (4 columns)
\toprule
TOI  & Telescope (inst.) & PI & Date       & Band & $\Delta m$@1\arcsec{} & $\rho$    & $\Delta m$ & PA     \\
     &                        &    &  &      & (mag)                 & ($\arcsec$) & (mag)      & (deg)   \\ 
\midrule
  TOI-1137 & Palomar (PHARO) & Ciardi &  & Kcont & 8.36 &  &  & \\
           & Lick (ShARCS) & Giacalone & 2019-10-15 & J  & 6.38&  &  & \\
           & Lick (ShARCS) & Giacalone & 2019-10-15 & Ks & 6.21 &  &  & \\
           & Gemini (Alopeke) & Howell & 2022-09-13 & 562nm & 7.02 &  &  & \\
           & Gemini (Alopeke) & Howell & 2022-09-13 & 832nm & 9.04 &  &  & \\ \midrule
  TOI-1837 & Palomar (PHARO) & Ciardi &  & J     & 8.91  &  &  & \\
           & Palomar (PHARO) & Ciardi &  & Hcont & 8.47  &  &  & \\
           & Palomar (PHARO) & Ciardi &  & Kcont & 7.88  &  &  & \\
           & SAI (Speckle) & Safonov & 2020-12-29 & 625 nm & 6.6 &$0.163\pm0.002$ & $3.03\pm 0.73$ & $76.4\pm0.3$\\
           & SAI (Speckle) & Safonov & 2021-01-23 & 625 nm & 7.2 & $0.150\pm0.003$ & $3.24 \pm 0.58$ & $78.5\pm0.2$\\
           & SAI (Speckle) & Safonov & 2021-07-15 & 625 nm & 6.9 & $0.154\pm0.002$ & $3.05 \pm 0.19$ & $76.52\pm0.1$\\
           & SAI (Speckle) & Safonov & 2026-03-03 & 625 nm & 7.5 & $0.116\pm0.001$ & $3.148 \pm 0.040$ & $75.8\pm0.4$\\
           & WIYIN (NESSI) & Everett & 2021-04-20 & 562nm& 6.02  & 0.16 & 3.38 & 77\\
           & WIYIN (NESSI) & Everett & 2021-04-20 & 832nm & 7.78 & 0.16 & 2.79 & 77\\
           & Lick (ShARCS) & Giacalone & 2020-11-30 & Ks & 4.68  &  &  & \\ \midrule
  TOI-2114 & Gemini (Alopeke) & Howell & 2024-05-26 & 562nm  & 6.20 &  &  & \\
           & Gemini (Alopeke) & Howell & 2024-05-26 &  832nm &  7.72  &  &  & \\
           & Lick (ShARCS) & Giacalone & 2021-05-31 &  &  &  &  & \\ 
           & Palomar (PHARO) & Ciardi &  & Br$\gamma$ & 8.33 &  &  & \\ \midrule
  TOI-4492 & Palomar (PHARO) & Ciardi &  & Hcont & 7.88 &  &  & \\
           & Palomar (PHARO) & Ciardi &  & Kcont & 8.11 &  &  & \\
           & Palomar (PHARO) & Ciardi &  & Br$\gamma$ & 8.21 &  &  & \\
           & WIYIN (NESSI) & Everett & 2022-04-18 & 832nm & 5.67 &  &  & \\
           & SAI (Speckle) & Safonov & 2021-10-30 & I-band & 7.11 &  &  & \\ \midrule
  TOI-5806 & Palomar (PHARO) & Ciardi &  & Hcont & 8.45 &  &  & \\
           & Palomar (PHARO) & Ciardi &  & Kcont &7.72  &  &  & \\
           & SAI (Speckle) & Safonov & 2022-11-20 & 625 nm & 6.7 & $2.098 \pm 0.020$ & $5.671 \pm 0.081$ & $381.9\pm0.4$\\
           & SAI (Speckle) & Safonov & 2023-07-04 & 625 nm & 6.3 & $2.100 \pm 0.020$ & $5.243 \pm 0.068$ & $381.4\pm0.4$\\
           & SOAR & Ziegler & 2022-11-04 & I-band & 5.04 & 2.10 & 4.1 (I-band) & \\
\bottomrule
\end{tabular}
\end{table*}

\begin{table*}
 \setlength{\extrarowheight}{3pt}
 \small
\caption{Radial velocity time series and corresponding corrections (drift and nightly zero points) applied for each of the seven targets analysed in this work. Only the first 5 lines of the table are shown. The full table is available online at the CDS under the following link \url{XXX}.}
\label{tab:RVdata}
\begin{tabular}{llllllllllll}
\toprule
TOI & Inst. & BJD    & RV      & $\sigma_{\rm RV}$ & drift & $\sigma_{\rm drift}$ & FWHM   & $\sigma_{\rm FWHM}$ & NZP   & $\sigma_{\rm NZP}$ & BERV$^{\dagger}$ \\
    &       & (days) & (m/s)   & (m/s)             & (m/s) & (m/s)                & (km/s) & (km/s)              & (m/s) & (m/s)              & (km/s) \\
\midrule
TOI-603 & ${\rm CAFE}_2$ & 2459528.6594786 & 31.7211 & 0.0070 & -0.2160 & 0.0030 & 11.5340 & 0.0061 & 63.9 & 1.9 & 29.9155 \\
TOI-603 & ${\rm CAFE}_2$ & 2459528.7243467 & 31.6659 & 0.0088 & -0.2150 & 0.0050 & 11.5960 & 0.0064 & 63.9 & 1.9 & 29.7895 \\
TOI-603 & ${\rm CAFE}_2$ & 2459529.6266098 & 31.7012 & 0.0075 & -0.1820 & 0.0030 & 11.5290 & 0.0065 & 10.0 & 2.2 & 29.9899 \\
TOI-603 & ${\rm CAFE}_2$ & 2459529.6879838 & 31.7855 & 0.0073 & -0.2200 & 0.0040 & 11.4490 & 0.0060 & 10.0 & 2.2 & 29.8872 \\
TOI-603 & ${\rm CAFE}_2$ & 2459530.628636 & 31.560 & 0.028 & -0.1310 & 0.0030 & 11.703 & 0.012 & -15.4 & 2.1 & 30.0034 \\
...     & ... & ...     & ...   & ...   & ...   & ...     & ...    & ...  & ... & ... \\
\bottomrule
\end{tabular}
\tablefoot{
\tablefoottext{$\dagger$}{Barycentric Earth Radial Velocity.}
}
\end{table*}
\FloatBarrier

\newpage
\onecolumn

\begin{landscape}
\setlength{\extrarowheight}{3pt}
\scriptsize
\setlength{\tabcolsep}{3pt}
\renewcommand{\arraystretch}{1.05}

\begin{longtable}{l *{10}{>{\raggedright\arraybackslash}p{0.10\textwidth}}}
\caption{Joint radial velocity and light-curve priors and posteriors for TOI-603, TOI-2114, TOI-4492, TOI-5806, and TOI-5811\,B.}
\label{tab:posteriors}\\

\toprule

& \multicolumn{2}{c}{TOI-603}
& \multicolumn{2}{c}{TOI-2114}
& \multicolumn{2}{c}{TOI-4492}
& \multicolumn{2}{c}{TOI-5806} 
& \multicolumn{2}{c}{TOI-5811\,B} \\
\\

\cmidrule(lr){2-3}\cmidrule(lr){4-5}\cmidrule(lr){6-7}\cmidrule(lr){8-9}\cmidrule(lr){10-11}

Parameter
& Prior & Posterior
& Prior & Posterior
& Prior & Posterior
& Prior & Posterior
& Prior & Posterior \\

\midrule

\endfirsthead

\toprule

& \multicolumn{2}{c}{TOI-603}
& \multicolumn{2}{c}{TOI-2114}
& \multicolumn{2}{c}{TOI-4492}
& \multicolumn{2}{c}{TOI-5806} 
& \multicolumn{2}{c}{TOI-5811\,B} \\

\cmidrule(lr){2-3}\cmidrule(lr){4-5}\cmidrule(lr){6-7}\cmidrule(lr){8-9}\cmidrule(lr){10-11}

Parameter
& Prior & Posterior
& Prior & Posterior
& Prior & Posterior
& Prior & Posterior
& Prior & Posterior \\

\midrule

\endhead

\midrule
\multicolumn{11}{r}{Continued on next page}\\

\endfoot

\bottomrule
\endlastfoot

% ==========================================================

% INFERRED PARAMETERS

% ==========================================================

\multicolumn{11}{l}{\textit{Inferred parameters}}\\

Orbital period, $P_b$ [days]
& $\mathcal{G}(\mu,\sigma)$ & $16.179846^{+0.000019}_{-0.000017}$
& $\mathcal{G}(\mu,\sigma)$ & $6.209959^{+0.0000045}_{-0.0000047}$
& $\mathcal{G}(\mu,\sigma)$ & $4.4331824^{+0.0000015}_{-0.0000014}$
& $\mathcal{G}(\mu,\sigma)$ & $3.1856381^{+0.0000010}_{-0.0000010}$ 
& $\mathcal{G}(\mu,\sigma)$ & $6.25663^{+0.00012}_{-0.00012}$\\

Time of mid-transit,
& $\mathcal{G}(\mu,\sigma)$ & $271.21509^{+0.00093}_{-0.00092}$
& $\mathcal{G}(\mu,\sigma)$ & $476.6557^{+0.0011}_{-0.0010}$
& $\mathcal{G}(\mu,\sigma)$ & $555.31972^{+0.00032}_{-0.00030}$
& $\mathcal{G}(\mu,\sigma)$ & $-149.18762^{+0.00015}_{-0.00016}$ 
& $\mathcal{G}(\mu,\sigma)$ & $549.422^{+0.013}_{-0.014}$ \\

~~~~~~($T_{0,b}-2460000$) [days]
&  &
&  &
&  &
&  & \\

RV semi-amplitude, $K_b$ [m/s]
& $\mathcal{U}(0,10000)$ & $8.0^{+1.4}_{-1.4}$
& $\mathcal{U}(0,10000)$ & $96.5^{+4.9}_{-4.5}$
& $\mathcal{U}(0,10000)$ & $712.8^{+6.7}_{-6.7}$
& $\mathcal{LU}(0.01,10000)$ & $237.1^{+9.4}_{-9.3}$ 
& $\mathcal{U}(0.0,10000.0)$ & $71.8^{+2.4}_{-2.5}$ \\

Orbital eccentricity, $e_b$
& -- & --
& $\mathcal{U}(0,0.8)$ & $0.472^{+0.026}_{-0.026}$
& -- & --
& -- & -- 
& -- & -- \\

Arg. periastron, $\omega_b$ [deg]
& -- & --
& $\mathcal{U}(0,360)$ & $211.2^{+5.5}_{-4.9}$
& -- & --
& -- & -- 
& -- & -- \\

Planet-star-radius, $R_{p,b}/R_\star$
& $\mathcal{LU}(10^{-4},10^0)$ & $0.04570^{+0.00031}_{-0.00037}$
& $\mathcal{LU}(10^{-4},10^0)$ & $0.06364^{+0.00046}_{-0.00044}$
& $\mathcal{LU}(10^{-4},10^0)$ & $0.08565^{+0.00097}_{-0.00079}$
& $\mathcal{LU}(10^{-4},10^0)$ & $0.0916^{+0.0062}_{-0.0111}$ 
& $\mathcal{LU}(10^{-4},0.0606)$ & $0.0540^{+0.0046}_{-0.0046}$ \\

Orbital inclination, $i_b$ [deg]
& $\mathcal{U}(50,90)$ & $89.58^{+0.39}_{-0.29}$
& $\mathcal{U}(70,90)$ & $84.39^{+0.52}_{-0.44}$
& $\mathcal{U}(70,90)$ & $88.78^{+0.84}_{-0.81}$
& $\mathcal{U}(70,90)$ & $82.24^{+0.39}_{-0.28}$ 
& $\mathcal{U}(70,90)$ & $76.72^{+0.72}_{-0.76}$ \\

Effective temperature, $T_{\rm eff}$ [K]
& $\mathcal{G}(6027,93)$ & $6027^{+96}_{-86}$
& $\mathcal{G}(6382,200)$ & $6380^{+200}_{-200}$
& $\mathcal{G}(5667,140)$ & $5660^{+140}_{-150}$
& $\mathcal{G}(6602,200)$ & $6610^{+200}_{-200}$ 
& $\mathcal{G}(4840,110)$ & $4879^{+85}_{-117}$ \\

Slope [m s$^{-1}$ day$^{-1}$]
& $\mathcal{U}(-5,5)$ & $0.023^{+0.060}_{-0.063}$
& $\mathcal{U}(-5,5)$ & $-0.06^{+0.078}_{-0.077}$
& $\mathcal{U}(-5,5)$ & $-0.028^{+0.049}_{-0.053}$
& $\mathcal{U}(-5,5)$ & $-0.25^{+0.26}_{-0.25}$ 
& -- & -- \\

$\eta_{\sigma}$ TESS [ppm]
& -- & --
& $\mathcal{U}(0,10)$ & $0.0187^{+0.0055}_{-0.0065}$
& $\mathcal{U}(0,10)$ & $1.747^{+0.043}_{-0.044}$
& $\mathcal{U}(0,10)$ & $0.0522^{+0.0038}_{-0.0038}$ 
& $\mathcal{U}(0,10)$ & $0.139^{+0.014}_{-0.017}$ \\

$\eta_{\rho}$ TESS [days]
& -- & --
& $\mathcal{U}(0.01,10)$ & $8.1^{+2.2}_{-1.4}$
& $\mathcal{U}(0.01,10)$ & $0.036785^{+4.9e-05}_{-4.4e-05}$
& $\mathcal{U}(0.01,10)$ & $0.0523^{+0.0085}_{-0.0092}$ 
& $\mathcal{U}(0.01,10)$ & $0.0141^{+0.0034}_{-0.0024}$\\

$\eta_1$ [m/s]
& -- & --
& -- & --
& -- & --
& $\mathcal{U}(-3,8)$ & $4.89^{+0.22}_{-0.28}$
& -- & -- \\

$\eta_2$ [days]
& -- & --
& -- & --
& -- & --
& $\mathcal{LU}(15,1000)$ & $48^{+10}_{-15}$ 
& -- & -- \\

$\eta_3$ [days]
& -- & --
& -- & --
& -- & --
& $\mathcal{U}(10,100)$ & $13.91^{+0.34}_{-0.31}$ 
& -- & -- \\

$\eta_4$
& -- & --
& -- & --
& -- & --
& $\mathcal{LU}(0.1,10)$ & $1.37^{+0.43}_{-0.70}$ 
& -- & -- \\

% ==========================================================

% DERIVED

% ==========================================================

\midrule

\multicolumn{11}{l}{\textit{Derived parameters}}\\

Planet mass
& derived & $33.0^{+6.5}_{-6.2}\,M_\oplus$
& derived & $1.02^{+0.15}_{-0.14}\,M_J$
& derived & $5.92^{+0.67}_{-0.64}\,M_J$
& derived & $2.27^{+0.26}_{-0.25}\,M_J$ 
& derived & $0.81^{+0.11}_{-0.10}\,M_J$ \\

Planet radius
& derived & $7.37^{+0.24}_{-0.23}\,R_\oplus$
& derived & $1.298^{+0.060}_{-0.060}\,R_J$
& derived & $1.051^{+0.055}_{-0.055}\,R_J$
& derived & $1.35^{+0.17}_{-0.11}\,R_J$ 
& derived & $1.96^{+0.18}_{-0.18}\,R_J$ \\

Planet density, $\rho_{b}$ [g$\cdot$~cm$^{-3}$]
& derived & $0.453^{+0.089}_{-0.084}$
& derived & $0.580^{+0.074}_{-0.072}$
& derived & $6.37^{+0.49}_{-0.56}$
& derived & $1.15^{+0.35}_{-0.39}$ 
& derived & $0.134^{+0.053}_{-0.035}$ \\

Orbit semi-major axis, $a_{b}$ [AU]
& derived & $0.1282^{+0.0046}_{-0.0049}$
& derived & $0.0760^{+0.0050}_{-0.0049}$
& derived & $0.0535^{+0.0029}_{-0.0030}$
& derived & $0.0484^{+0.0025}_{-0.0025}$ 
& derived & $0.0731^{+0.0048}_{-0.0047}$ \\

Relative orbital separation $a_b/R_\star$
& derived & $18.76^{+0.20}_{-0.48}$
& derived & $7.8^{+0.35}_{-0.37}$
& derived & $9.17^{+0.15}_{-0.25}$
& derived & $6.91^{+0.17}_{-0.21}$
& derived & $4.22^{+0.20}_{-0.20}$ \\

Transit duration [h]
& derived & $6.834^{+0.043}_{-0.037}$
& derived & $4.515^{+0.095}_{-0.092}$
& derived & $3.952^{+0.025}_{-0.023}$
& derived & $2.016^{+0.035}_{-0.027}$
& derived & $4.85^{+0.17}_{-0.15}$ \\

Impact parameter
& derived & $0.137^{+0.121}_{-0.095}$
& derived & $0.763^{+0.034}_{-0.029}$
& derived & $0.20^{+0.12}_{-0.13}$
& derived & $0.933^{+0.018}_{-0.011}$
& derived & $0.9690^{+0.0091}_{-0.0107}$ \\

Incident Flux, $F_{\rm inc,b}$ [$F_{{\rm inc},\oplus}$]
& derived & $157^{+12}_{-11}$
& derived & $1130^{+190}_{-160}$
& derived & $510^{+61}_{-55}$
& derived & $1660^{+240}_{-200}$
& derived & $1320^{+180}_{-180}$ \\

Stellar luminosity, $L_{\star}$ [$L_{\odot}$]
& derived & $2.57^{+0.23}_{-0.22}$
& derived & $6.5^{+1.08}_{-0.95}$
& derived & $1.46^{+0.22}_{-0.20}$
& derived & $3.89^{+0.62}_{-0.54}$
& derived & $7.01^{+0.90}_{-0.86}$ \\

Equilibrium temperature, $T_{\rm eq,b}$ [K]
& derived & $899^{+17}_{-17}$
& derived & $1472^{+57}_{-55}$
& derived & $1207^{+34}_{-34}$
& derived & $1621^{+56}_{-52}$ 
& derived & $1530^{+50}_{-54}$ \\

\end{longtable}
\tablefoot{
\tablefoottext{$\dagger$}{The Gaussian priors for the orbital period and time of conjunction for each of the targets are as follows. TOI-603: $\mathcal{G}(16.1798603,0.0000461)$, $\mathcal{G}(60271.218286,0.0023676)$; TOI-2114: $\mathcal{G}(6.2099619,0.00006)$, $\mathcal{G}(60476.661598,0.009895)$; TOI-4492: $\mathcal{G}(4.4331838,0.000027)$, $\mathcal{G}(60555.320385,0.005313)$; TOI-5806: $\mathcal{G}(3.1856488,0.000391)$, $\mathcal{G}(59850.812572,0.002427)$; ; TOI-5811\,B: $\mathcal{G}(6.25656,0.000276)$, $\mathcal{G}(60549.411097,0.021504)$.}
}

\end{landscape}

\end{appendix}

\end{document}